\documentclass[12pt,a4paper]{article}

\usepackage[english]{babel}
\usepackage{pdfpages}
\usepackage[utf8x]{inputenc}
\usepackage[T1]{fontenc}
\usepackage[a4paper,top=2cm,bottom=3cm,left=2cm,right=2cm,marginparwidth= 2cm]{geometry}
\usepackage{pgfplots}
\pgfplotsset{compat=1.18}
\usepgfplotslibrary{external}
\usepackage{tikz}
\usepackage{amsmath,amsthm,amssymb,mathrsfs}
\usepackage{graphicx}
\graphicspath{{images/}}
\usepackage{url}
\usepackage[colorinlistoftodos]{todonotes}
\usepackage[colorlinks=true, allcolors=blue]{hyperref}
\usepackage{dsfont}
\usepackage[dvipsnames]{xcolor}
\usepackage[normalem]{ulem} 

\theoremstyle{plain}

\theoremstyle{definition}

\newtheorem{rem}{Remark}

\newcommand{\be}{\begin{equation}}
\newcommand{\ee}{\end{equation}}

\renewcommand{\Re}{\mathrm{Re}}
\renewcommand{\Im}{\mathrm{Im}}

\newcommand{\CCC}{\mathbb{C}}

\newcommand{\HHH}{\mathbb{H}}
\newcommand{\MMM}{\mathbb{M}}

\newcommand{\PPP}{\mathbb{P}}
\newcommand{\RRR}{\mathbb{R}}

\newcommand{\va}{\boldsymbol{a}}
\newcommand{\vq}{\boldsymbol{q}}
\newcommand{\vb}{\boldsymbol{b}}
\newcommand{\vj}{\boldsymbol{j}}
\newcommand{\vk}{\boldsymbol{k}}
\newcommand{\vn}{\boldsymbol{n}}
\newcommand{\vp}{\boldsymbol{p}}
\newcommand{\vx}{\boldsymbol{x}}
\newcommand{\vu}{\boldsymbol{u}}
\newcommand{\vv}{\boldsymbol{v}}
\newcommand{\vA}{\boldsymbol{A}}
\newcommand{\vK}{\boldsymbol{K}}
\newcommand{\vQ}{\boldsymbol{Q}}
\newcommand{\vX}{\boldsymbol{X}}
\newcommand{\valpha}{\boldsymbol{\alpha}}
\newcommand{\vsigma}{\boldsymbol{\sigma}}
\newcommand{\vzero}{\boldsymbol{0}}

\newcommand{\nslash}{n\!\!\!/}
\newcommand{\timestep}{\mathscr{T}}
\newcommand{\timescaled}{\tau}
\newcommand{\inc}{\Psi_{\text{inc}}}
\newcommand{\refl}{\Psi_{\text{ref}}}
\newcommand{\trans}{\Psi_{\text{trans}}}

\begin{document}
\title{\textbf{Scattering Cross Section Formula Derived From Macroscopic Model of Detectors}}

\author{Rashi Kaimal\footnote{Fachbereich Mathematik, Eberhard-Karls-Universit\"at T\"ubingen, Auf der Morgenstelle 10, 72076 T\"ubingen, Germany. Email: rashi-ramesh.kaimal@student.uni-tuebingen.de}~~and Roderich Tumulka\footnote{Fachbereich Mathematik, Eberhard-Karls-Universit\"at T\"ubingen, Auf der Morgenstelle 10, 72076 T\"ubingen, Germany. Email: roderich.tumulka@uni-tuebingen.de}}
\date{June 8, 2026}

\maketitle

\begin{abstract}
We are concerned with the justification of the statement, commonly (explicitly or implicitly) used in quantum scattering theory, that for a free non-relativistic quantum particle with initial wave function $\Psi_0(\vx)$, surrounded by detectors along a sphere of large radius $R$, the probability distribution of the detection time and place has asymptotic density (i.e., scattering cross section) $\sigma(\vx,t)= m^3 \hbar^{-3} R t^{-4} |\widehat{\Psi}_0(m\vx/\hbar t)|^2$ with $\widehat{\Psi}_0$ the Fourier transform of $\Psi_0$. We give two derivations of this formula, based on different macroscopic models of the detection process. The first one consists of a negative imaginary potential of strength $\lambda>0$ in the detector volume (i.e., outside the sphere of radius $R$) in the limit $R\to\infty,\lambda\to 0, R\lambda\to \infty$. The second one consists of repeated nearly-projective measurements of (approximately) the observable $1_{|\vx|>R}$ at times $\timestep,2\timestep,3\timestep,\ldots$ in the limit $R\to\infty,\timestep\to\infty,\timestep/R\to 0$; this setup is similar to that of the quantum Zeno effect, except that there one considers $\timestep\to 0$ instead of $\timestep\to\infty$. We also provide a comparison to Bohmian mechanics: while in the absence of detectors, the arrival times and places of the Bohmian trajectories on the sphere of radius $R$ have asymptotic distribution density given by the same formula as $\sigma$, their deviation from the detection times and places is not necessarily small, although it is small compared to $R$, so the effect of the presence of detectors on the particle can be neglected in the far-field regime. We also cover the generalization to surfaces with non-spherical shape, to the case of $N$ non-interacting particles, to time-dependent surfaces, and to the Dirac equation.

\bigskip

\end{abstract}

\section{Introduction}
\label{sec:intro}

We are concerned with the formula for the scattering cross section in non-relativistic 1-particle quantum mechanics (e.g., \cite{AJS77,New82,Daumer,DGMZ06}),
\be\label{cross}
\sigma(\vx,t)= \frac{m^3 R}{\hbar^3 t^4} \, \Bigl|\widehat{\Psi}_0\Bigl(\frac{m\vx}{\hbar t}\Bigr)\Bigr|^2 \,.
\ee
Here, one considers detectors placed around the particle along a sphere $S_R=\{\vx\in\RRR^3: |\vx|=R\}$ of large radius $R$ (far-field regime) around the origin in $\RRR^3$, $\sigma(\vx,t) \, d^2\vx \, dt$ means the probability that detectors click in the time interval $[t,t+dt]$ in a surface element of area $d^2\vx$ around $\vx$, and $\widehat{\Psi}_0$ is the Fourier transform of the initial wave function $\Psi_0$. (For generalizations to other shapes than spheres, more particles, time-dependent surfaces, and the relativistic case, see Remarks~\ref{rem:shape}--\ref{rem:Dirac} and Section~\ref{sec:general}.) It is assumed for formula \eqref{cross} that the particle does not interact with anything after time $t=0$ except the detectors, so any interaction with a target near the origin (such as a nucleus) is taken to be completed at time zero, to which the wave function $\Psi_0$ refers that we take here as the ``initial'' wave function. That is, we focus here on the last stage or ``free phase'' of the evolution; for an analysis of the prior stages, see \cite{DGMZ06,Kom12}. In particular, whenever it is appropriate to apply an $S$-matrix, then we assume that $\Psi_0$ is the wave function after the application of the $S$-matrix.

The formula \eqref{cross}, which agrees with empirical data (for further discussion see \cite{AJS77, Coplan1994,Ullrich2003}), is usually theoretically justified by arguing that $\int_{B_R} |\Psi_t(\vx)|^2 \, d^3\vx$ (where $B_R=\{\vx\in\RRR^3: |\vx|<R\}$ denotes the ball of radius $R$ around the origin) should be the probability that no detection has occurred by time $t>0$, and that this quantity changes in $[t,t+dt]$ (to leading order in $R$) by $-dt \int_{S_R} \sigma(\vx,t) \, d^2\vx$ as given by \eqref{cross}. As we will discuss in Section~\ref{Comparisons}, this justification has some merits but is not satisfactory. 
Our goal in this paper is to provide a better justification of \eqref{cross} by deriving that it is, not just \emph{a desirable} probability density, but \emph{the actual} probability density of when and where detectors click. Our derivation starts from a \emph{model} of detectors.
We present derivations from two such models, the first using imaginary potentials and the second repeated (``stroboscopic'') nearly-projective position measurements outside $B_R$ as effective macroscopic (and approximate) descriptions of detectors. It would be of interest to have a derivation starting from a \emph{microscopic} model of the detectors as quantum-mechanical many-body systems, but this will not be attempted here; however, an outline of such a derivation can be obtained by combining the reasoning in this paper with the outlined derivations in \cite{Hal99,detect-derive} of imaginary potentials from microscopic detector models.  

We are not aware of any works deriving \eqref{cross} from any model of detectors except the current work in progress of Beck and Lazarovici \cite{BL25}. In another context, macroscopic models of detectors were used in \cite{LT20,LT22} for detection on spacelike hypersurfaces, including a ``stroboscopic'' one in \cite{LT20}.

One aspect of particular interest is the study of the effect of the presence of detectors as a disturbance on the particle or its wave function \cite{GTZ24}. This effect can be calculated only if the detectors are modeled in some way. In fact, we obtain the size of these disturbances as a subleading correction to \eqref{cross} in the appropriate limit; in this limit, $R\to\infty$ while the strength $\lambda$ of the imaginary potential tends to 0 with $\lambda R\to\infty$, or the time difference $\timestep$ of the nearly-projective position measurements tends to $\infty$ with $\timestep/R\to 0$. (Note that the quantum Zeno effect \cite{Zeno} would occur in the limit $\timestep\to0$.) 

Specifically in Bohmian mechanics (e.g., \cite{DT09,DGZ13,Tum22}), one can consider the disturbance caused by the detectors in the following way. In Bohmian mechanics, quantum particles have trajectories and thus a well-defined arrival time $T_{WOD}$ in the absence of detectors (WOD = without detectors).  One can now ask, how well the time $T_D$ of detection approximates $T_{WOD}$. One can also compare both $T_D$ and $T_{WOD}$ to the time $T_{WID}$ at which the trajectory crosses the sphere $S_R$ in the presence of detectors (WID = with detectors; see \cite{GTZ24} for discussion). One may expect that the particle gets detected at the time it arrives (i.e., $T_D\approx T_{WID}$), but still the question arises how accurately this is true. We find, for the model involving imaginary potentials as well as the one with repeated position measurements, that in the limit mentioned above,
\be\label{TDTWIDTWOD}
T_D-T_{WID}~\longrightarrow~ \infty~~~\text{and}~~~T_{WID}- T_{WOD} ~\longrightarrow~ \infty
\ee
while
\be\label{TDTWIDTWODR}
\frac{T_{D}-T_{WID}}{R} ~\longrightarrow~ 0~~~\text{and}~~~\frac{T_{WID}- T_{WOD}}{R} ~\longrightarrow~ 0 \,.
\ee
Therefore, the corrections are negligible compared to $R$ (note that $T_{D}$ will be of the same order of magnitude as $R$) but not on the absolute scale. That is, as long as we consider quantities only on the scale of $R$, it is justified to equate $T_D$, $T_{WID}$, and $T_{WOD}$. At the same time, our consideration illustrates the conceptual difference between the three quantities, and makes clear that they should not be expected to agree outside the scattering regime or on the microscopic scale.

The comparison of $T_D,T_{WID}$, and $T_{WOD}$ is based on a comparison of the evolution of the wave function $\Psi_{WID}$ in our models with the free evolution (i.e., with $V=0$) of the wave function $\Psi_{WOD}$ without detectors for large times $t=\mathcal{O}(R)$ and distances $|\vx|=\mathcal{O}(R)$, which takes the following well-known asymptotic form, as derived in, e.g., \cite[Prop.~3.17]{AJS77}, \cite[Thm.~IX.31]{RS2}, \cite[(9.20)]{DT09}: 
\begin{equation}\label{asymppsi}
    \Psi_{WOD}(\vx,t)\approx\ \left(\frac{m}{i\hbar t}\right)^{3/2} \widehat{\Psi}_0(\vk) \, \exp(i\vk\cdot \vx) \, \exp\left(-i\omega t\right)~~~\text{with}~~\vk=\frac{m\vx}{\hbar t},~\omega=\frac{\hbar \vk^2}{2m}.
\end{equation}
This formula expresses that $\Psi_{WOD}(\vx,t)$ is locally a plane wave $\exp(i\vk\cdot \vx -i\omega t)$ whose amplitude $(m/i\hbar t)^{3/2} \widehat{\Psi}_0(\vk)$ varies slowly with $\vx$ and $t$. It also expresses that in the long run, the Fourier modes of $\Psi_0$ get separated in space as they move at different velocities given by $\hbar \vk/m$.
The error in this approximation is of the order $1/R$.

Let us now leave aside the estimates concerning $T_{WID}$ and $T_{WOD}$, which are defined only in Bohmian mechanics, and turn to our main results, which concern $T_{D}$ and $\vX_{D}$ and are independent of the Bohmian framework. We obtain for both models of the detection process (imaginary potentials and repeated position measurements) that the leading contribution (i.e., on the scale of $R$) to the probability distribution of the time $T_{D}$ and the place $\vX_{D}$ of detection (also called the ``screen observables'' or ``detection observables'') in the relevant limit is given by \eqref{cross}. Put differently, for any surface element $d^2\vu$ of the unit sphere and any intervals $d\rho$ (of distance rescaled by $R^{-1}$) and $d\timescaled$ (of time rescaled by $R^{-1}$),
\be\label{cross3}
\mathbb{P}\biggl(\frac{\vX_{D}}{|\vX_D|}\in d^2\vu,~ \frac{|\vX_D|}{R} \in d\rho,~  \frac{T_{D}}{R}\in d\timescaled\biggr) ~~\longrightarrow~~ \frac{m^3}{\hbar^3 \timescaled^4} \,  \Bigl|\widehat{\Psi}_0\Bigl(\frac{m\vu}{\hbar \timescaled}\Bigr)\Bigr|^2 \, \delta(\rho-1)\, d^2\vu \, d\rho\,  d\timescaled
\ee
in the relevant limits involving $R$ and $\lambda$ or $\timestep$.
(We write $d^2\vu$ for both the infinitesimal set and its area.)
This entails that neglecting the effect of the presence of detectors is justifiable in the scattering regime. Another equivalent way of expressing the joint distribution \eqref{cross3} of $\vX_D/R$ and $T_D/R$ in the limit $R\to\infty$ is to say that they are outcomes of ideal quantum measurements of the commuting observables 
\be\label{observables}
|\hat{\vp}|^{-1}\hat{\vp}~~\text{and}~~m|\hat{\vp}|^{-1} \,,
\ee
where $\hat{\vp}=(\hat{p}_1,\hat{p}_2,\hat{p}_3)$ is the triple of momentum operators.

\begin{rem}
A basic difficulty about \emph{deriving} the scattering cross section \eqref{cross} arises from the question what to derive it \emph{from}: one first needs to have a description of detectors that is generally applicable. And this is controversial: not only are there many inequivalent proposals for how to compute the probability distribution of the detection time (e.g., \cite{AB61,All69b,ML00}), but also even arguments suggesting the impossibility of such a distribution, such as (i)~Pauli's \cite[p.~140]{Pauli} argument that there cannot be any self-adjoint time operator (canonically conjugate to the Hamiltonian) and (ii)~the quantum Zeno effect, which was formulated by some researchers as leading to the absurd conclusion that ``a watched pot never boils'' \cite{Zeno}. Our answer to the question about the starting point of the derivation is to use two models of detection (imaginary potentials and repeated position measurements) that are contradiction-free and lead, as we show, to a non-trivial distribution in agreement with \eqref{cross}.
\hfill$\diamond$\end{rem}

\begin{rem}\label{rem:hard}
Another difficulty arises because one might expect that the detectors should be modeled as a detecting surface at $S_R$, that is, as \emph{hard} detectors (ones that detect the particle immediately upon arrival, as opposed to \emph{soft} detectors which may take a while to notice the particle). Hard detectors lead not only to the aforementioned problems of controversial, inequivalent proposals and paradoxes like the quantum Zeno effect, but also, even in paradox-free models of hard detectors such as the absorbing boundary rule \cite{detect-rule}, to reflection of part of the wave function from the detector surface, which is undesirable and would not fit together with \eqref{cross}. Our answer to this difficulty is to allow soft (but not too soft) detectors for our derivation, as suggested already in \cite[Sec.~5]{detect-rule}; we have in mind the following reasons:
\begin{enumerate}
    \item[(i)] One wants to avoid reflection of a significant amount of wave from the detector (for example for the reasons given in Section~\ref{sec:common}), but this would be inevitable for a hard detector, see Section~\ref{sec:ABR}. Recently, Cavendish and Das \cite{cavendish2025absorbing} have given a quantitative analysis of how the absorbing boundary rule deviates from \eqref{cross}.
    \item[(ii)] We can afford some softness because $T_D$ is large like $R$, so if we desire a small relative error then the absolute error can still be large, as long as it is small compared to $R$. 
    \item[(iii)]  In practice, the detectors used in (e.g.) particle accelerators are quite soft. It would be interesting to see quantitative measures of their degree of softness or hardness, but we do not do that here.
\end{enumerate}

In our treatment, the softness is reflected in the use of imaginary potentials of strength $\lambda \to 0$ and, in the model of repeated position measurements, in the use of large time intervals $\timestep$ between these measurements as well as in the use of an approximation of the function
\be\label{1def}
1_{|\vx|>R} = \begin{cases}
    1 & \text{if }|\vx|>R\\
    0 & \text{otherwise}
\end{cases}
\ee
by a continuous function. In more detail, the expression \eqref{1def} can be regarded as a function of the variable $\vx$ or as a self-adjoint operator (the corresponding multiplication operator). Since the function assumes only the values 0 and 1, the operator has only eigenvalues 0 and 1, so it is a projection operator: for any $\psi$, it yields the part of $\psi$ outside the ball $B_R$. To measure this observable means intuitively to check whether the particle is outside or inside $B_R$. This \emph{projective measurement} will be replaced by a \emph{nearly-projective measurement} in order to allow for soft detectors, as described in Section~\ref{sec:Zenoresults}. 
\hfill$\diamond$\end{rem}

\begin{rem}\label{rem:shape}
    Our results generalize to other shapes of the region $\Omega\subset \RRR^3$ enclosed by the detectors instead of balls (e.g., to cubes or ellipsoids) as follows, provided $\Omega$ is \textit{star-shaped} (see \eqref{concavesurf} for the definition). We describe the surface $\partial \Omega$ of $\Omega$ by means of a function $s(\vu)$ for every unit vector $\vu\in\RRR^3$ giving the distance of the surface from the origin in the direction $\vu$, and then scale by the factor $R$, so
    \be\label{s}
    \partial\Omega = \bigl\{\vx\in\RRR^3: |\vx|=Rs(\vx/|\vx|)\bigr\}\,.
    \ee
    Let $\vn(\vx)$ denote the outward unit normal vector to $\partial\Omega$ at $\vx\in\partial\Omega$ (which can be computed from the $s$ function and $R$). We assume that 
    \be\label{nxx+ve}
    \vn(\vx)\cdot \vx >0
    \ee 
    for $\vx\in\partial\Omega$. The formula analogous to \eqref{cross} then reads
    \be\label{crossn}
    \sigma(\vx,t)= \frac{m^3 \vn(\vx) \cdot \vx}{\hbar^3 t^4} \, \Bigl|\widehat{\Psi}_0\Bigl(\frac{m\vx}{\hbar t}\Bigr)\Bigr|^2 ~~~\text{for }\vx\in\partial\Omega\ ,
    \ee
    and will be justified in Section~\ref{sec:general}. 
\hfill$\diamond$\end{rem}

\begin{rem}\label{rem:N}
    Another generalization concerns the case of $N$ non-interacting (but entangled) particles. The formula analogous to \eqref{crossn} then reads
    \be\label{crossnN}
    \sigma(\vx_1,t_1,\ldots,\vx_N,t_N) = \frac{1}{\hbar^{3N}}  \Bigl( \prod_{i=1}^N \frac{m_i^3 \vn(\vx_i) \cdot \vx_i}{t_i^4} \Bigr) \, \Bigl|\widehat{\Psi}_0\Bigl(\frac{m_1\vx_1}{\hbar t_1},\ldots,\frac{m_N\vx_N}{\hbar t_N}\Bigr)\Bigr|^2 \,,
    \ee
    where $\sigma(\vx_1,t_1,\ldots,\vx_N,t_N) \, d^2\vx_1 dt_1 \cdots d^2\vx_N dt_N$ means the probability that each particle $i=1,\ldots,N$ will be detected in the surface element $d^2\vx_i$ and time interval $dt_i$ (related formulas are given in \cite[Eq.s (33) and (35)]{DT04}).
    This formula will also be justified in Section~\ref{sec:general}.
\hfill$\diamond$\end{rem}

\begin{rem}\label{rem:Dirac}
Further generalizations concern the Dirac equation instead of the non-relativistic Schr\"odinger equation and a time-dependent detecting surface. To this end, let $\MMM=\RRR^4$ be Minkowski space-time, $\MMM_{\geq}=\{x^0\geq 0\}$ in a certain Lorentz frame, and ${}^4\Omega\subseteq \MMM_{\geq}$ the space-time region surrounded by detectors at the surface $\partial_{>}{}^4\Omega:= \partial({}^4\Omega)\setminus \{x^0=0\}$. We assume that ${}^4 \Omega$ is obtained through scaling by the factor $R$ from a fixed set ${}^4\Omega_1$, and that the normal direction to $\partial_{>}{}^4\Omega$ is spacelike or timelike almost everywhere (that is, lightlike only on a null set of points); we define $n(x)=n^\mu=(n^0,\vn)$ to be the outer unit normal 4-vector at $x=x^\mu=(x^0,\vx)=(ct,\vx)\in\partial_{>}{}^4\Omega$ and assume that $n_\mu(x) x^\mu >0$ for timelike $x$. For $i=1,\ldots,N$, let $X_{D,i}=(cT_{D,i},\vX_{D,i})\in \partial_{>}{}^4\Omega$ be the point where particle $i$ gets detected. The scattering cross section $\sigma$ is then defined by the leading order term (as $R\to\infty$) of the relation
\be\label{sigmareldef}
\PPP\Bigl( X_{D,1}\in d^3x_1, \ldots, X_{D,N}\in d^3 x_N \Bigr) = \sigma(x_1,\ldots, x_N) \, c^{-N}V_3(d^3x_1)\cdots V_3(d^3x_N)\,,
\ee
where $d^3x$ means a surface element in $\partial_{>}{}^4\Omega$ around $x$ and $V_3(d^3x)$ its invariant 3-volume (= $|\det ({}^3g)|^{1/2} d^3x$ in coordinates). Let $\Psi_0$ be the ($\CCC^4$-valued) wave function on the initial surface $\{x^0=0\}$. If $\Psi_0$ involves only contributions with positive energy, then the formula analogous to \eqref{crossnN} reads
\be\label{crossDirac}
\sigma(x_1,\ldots,x_N) = \begin{cases}
0 ~~\text{if any $x_i$ is spacelike or lightlike, else}\\
\frac{c^{4N}}{\hbar^{3N}} \biggl( \prod\limits_{i=1}^N \frac{m_i^3 x_i^0 [n_\mu(x_i) x_i^\mu]}{|x_i|^5} \biggr) \biggl| \widehat{\Psi}_0 \biggl( \frac{m_1 c \vx_1}{\hbar |x_1|}, \ldots, \frac{m_N c \vx_N}{\hbar |x_N|} \biggr) \biggr|^2 \,,
\end{cases}
\ee
where $m_i>0$ is the mass of particle $i$, $\widehat{\Psi}_0$ the $3N$-dimensional Fourier transform of $\Psi_0$ (as before), and $|x|=\sqrt{x^\mu x_\mu}$ is the Minkowski length of a timelike 4-vector $x$. We derive this formula from imaginary potentials (but not the Zeno dynamics) in Section~\ref{sec:Dirac}, where we also provide the asymptotics of the wave function analogous to \eqref{asymppsi} also with contributions of negative energy, as well as the asymptotic form of the Bohmian trajectories (which are not straight lines but helices, featuring what can be regarded as a Bohmian analog of ``Zitterbewegung''). By the way, \eqref{crossDirac} implies a \emph{no-signaling theorem} for the scattering regime, see Section~\ref{sec:nosignaling}.
\hfill$\diamond$\end{rem}

\bigskip

The remainder of this paper is structured as follows. 
In the following Section \ref{Comparisons}, we review some common justifications for the scattering cross section formula and how our approach improves on them.
We also discuss the absorbing boundary rule and explain why it is not used to model the detectors here. We then summarize our key results in Section \ref{Results}. 
Sections \ref{Impot} and \ref{Zeno} provide the detailed derivations of the results presented in Section \ref{Results}. Finally, in Section~\ref{sec:general}, we generalize our results to other shapes, several particles, and the Dirac equation.

\section{Comparisons}
\label{Comparisons}

We now compare our way of arriving at the scattering cross section formula \eqref{cross} to ways that were previously described in the literature.

\subsection{Common Justifications}
\label{sec:common}

{\it Transport of probability-} As mentioned in the introduction, a common justification of \eqref{cross} is that $\int_{B_R} |\Psi_t(\vx)|^2 \, d^3\vx$, which expresses the amount of probability contained in $B_R$ at time $t$, should agree with the probability that no detection has occurred by $t$, or $\PPP(T_D>t)$. From this it follows by the continuity equation
\be
\frac{\partial}{\partial t}|\Psi|^2 = - \nabla \cdot \vj
\ee
with 
\be\label{jdef}
\vj(\vx,t) = \tfrac{\hbar}{m} \Im\bigl[\Psi(\vx,t)^* \, \nabla \Psi(\vx,t)\bigr]
\ee
the probability current vector field 
together with the Gauss integral theorem that 
\be\label{PTDj}
\PPP\bigl(T_D\in[t,t+dt]\bigr) = dt \int_{S_R} \frac{\vx}{R}\cdot \vj(x,t) \, d^2\vx \,.
\ee

There is another reasoning leading more directly to \eqref{PTDj}: It seems plausible that $\PPP(T_D\in[t,t+dt])$ should equal the amount of probability transported across $S_R$ during $[t,t+dt]$. Given that probability gets transported according to the current $\vj(\vx,t)$, the latter quantity equals the right-hand side of \eqref{PTDj}.

Inserting the asymptotic form
\eqref{asymppsi} of $\Psi$ into \eqref{jdef} and the result into \eqref{PTDj}, we find to leading order that
\be\label{PTDhatPsi}
\PPP\bigl(T_D\in[t,t+dt]\bigr) = dt  \frac{m^3R}{\hbar^3 t^4} \int_{S_R} \Bigl|\widehat{\Psi}_0\Bigl( \frac{m\vx}{\hbar t} \Bigr)\Bigr|^2 \, d^2\vx \,,
\ee
which is just \eqref{cross} integrated over $\vx\in S_R$. A similar joint consideration of $\vX_D$ and $T_D$ leads to \eqref{cross}. A reasoning of this kind is given in, e.g., \cite[Sec.~6.7]{New82}.

While the result \eqref{cross} is correct, we argue that the reasoning is not. A first way of seeing this is to notice that the largeness of $R$ and the spherical shape of the detector surface played a role in going from \eqref{PTDj} to \eqref{PTDhatPsi}, but not in the derivation of \eqref{PTDj}; thus, if that derivation was valid, it should apply to arbitrary surfaces instead of large spheres. But for arbitrary surfaces, the right-hand side of \eqref{PTDj} may be negative, so \eqref{PTDj} cannot be generally true. Moreover, it was shown in \cite{VHD13} that even if the right-hand side of \eqref{PTDj} is positive, it is in general not given by a POVM and thus cannot be the distribution of the outcome of any experiment \cite{DGZ04}, \cite[Sec.~5.1.2]{Tum22}.

Secondly, the more fundamental flaw with the reasoning outlined above is that it does not take into account the effect of the presence of detectors on the wave function. For example, placing a detector in one slit of a 2-slit experiment will famously destroy the interference, but the reasoning behind \eqref{PTDj} contains no consideration of such an effect, and indeed \eqref{PTDj}, when applied literally to a surface containing (e.g.) the left slit, the right half of the screen, and the plane perpendicular to the screen in the middle between the slits, will fail to predict the disappearance of the interference pattern. Now in the scattering regime, the effect of the presence of detectors is negligible, and to show this is our goal.

Some authors (e.g., \cite{MS20}) take for granted that $|\Psi_{WOD}(\vx,t)|^2$ as a function of $t$ for fixed $\vx$ is proportional to the probability density of the detection time if a detector is located at $\vx$. It is also tempting to compute the probability that a detector placed at location $\vx$ at time $t$ will click and take it, for fixed $\vx$, as the probability density at $t$ of the detection time. However, these identifications are not necessarily warranted.

\bigskip

{\it Classical motion-} Aharonov and Bohm \cite{AB61} had suggested to compute the distribution of the time of detection by first finding a formula for the analogous arrival time in classical mechanics, then quantizing this formula to obtain a self-adjoint observable, and then to determine the Born distribution of that observable. In our case, a classical particle starting at position $\vx_0$ with momentum $\vp_0$ would arrive on $S_R$ at time $Rm/|\vp_0|$ and place $R\vp_0/|\vp_0|$ (to leading order as $R\to\infty$), whose quantization (after rescaling by $R^{-1}$) is given by \eqref{observables}.

This line of reasoning gets reinforced by two observations: 
First, the plane wave formula \eqref{asymppsi} has the consequence that $|\Psi(\vx,t)|^2 \approx (m/\hbar t)^{3} |\widehat{\Psi}_0(m\vx/\hbar t)|^2$, which is the same large-time position distribution as that of a classical particle with momentum distribution $\rho_p(\vp)=\hbar^{-3} \,|\widehat{\Psi}_0(\vp/\hbar)|^2$, viz., $\rho_x(\vx,t) = (m/t)^3\rho_p(m\vx/t)$. For such a classical particle, the distribution of arrival time and place on $S_R$ has distribution density $\sigma_{\mathrm{class}}(\vx,t)=(m^3R/t^4)\,\rho_p(m\vx/t)$ to leading order in large $R$, which leads to \eqref{cross}. A reasoning of this kind for $\vX_D$ is given in \cite[Sec.~3-3]{AJS77}. Second, according to \eqref{asymppsi}, $\Psi_t$ is locally a plane wave, each of the local waves comes from a certain Fourier mode of $\Psi_0$, and they move at different velocities $\vv$ according to the de Broglie relation $m\vv=\hbar \vk$. This suggests that the mode with wave vector $\vk$ crosses $S_R$ for large $R$ at time $t=Rm/\hbar |\vk|$ in a neighborhood of $\vx=R\vk/|\vk|$.

The last two ways of reasoning are guesses with some plausibility, but we ask in this paper whether these guesses can be confirmed by a derivation. The argument of Aharonov and Bohm is even more dubious because a plausible generalization to arbitrarily shaped surfaces at arbitrary distances is lacking. On top of that, again none of these considerations takes the effect of the presence of detectors on the wave function into account.

\bigskip

{\it Bohmian motion-} Since in Bohmian mechanics, a quantum particle has a trajectory, it has a well defined time and place of first arrival on $S_R$ in the absence of detectors, $T_{WOD}$ and $\vX_{WOD}$, and it turns out that their joint probability distribution is given for large $R$ by \eqref{cross}. This fact has been presented \cite{Daumer,DT04,DGMZ06} as a justification of \eqref{cross} for $T_D$ and $\vX_D$. This argument leaves to be desired because first, it leaves open whether, how, and to which extent the presence of detectors might affect the wave function and therefore the trajectories and arrival time and place, and second, whether (e.g.)\ inefficiency of the detector might lead to deviations of the time $T_D$ of detection from the time $T_{WID}$ the Bohmian trajectory crossed $S_R$ in the presence of detectors \cite{GTZ24}. In fact, we provide results on the magnitude of the differences between $T_{WOD},T_{WID}$, and $T_D$. Moreover, it has been shown \cite{VHD13,GTZ23} that for general surfaces that are not far away, the distribution of $T_{WOD}$ is not given by a POVM, so there is no experiment whose outcome is $T_{WOD}$, so $T_{WOD} \neq T_D$.

\bigskip

In sum, even though these approaches lead to distributions of the detection time that agree with empirical observations in the scattering regime, our approach provides more, as it provides a derivation of what detectors will see, starting from some simple models of the detection mechanism.
All of these previous treatments take for granted that the detectors are ``passive'' and do not influence the particle under observation. But this cannot be taken for granted, and will not be taken for granted here.

\subsection{Absorbing Boundary Rule}
\label{sec:ABR}

The absorbing boundary rule (ABR) \cite{detect-rule} (see also \cite{Wer87,Fro25}) is a promising candidate for the joint distribution of $T_D$ and $\vX_D$ found by hard detectors along an arbitrary surface. It asserts that if a quantum particle in a region $\Omega$ is surrounded by detectors along the surface $\partial \Omega$, then the joint distribution of detection time and place can be obtained by solving a 1-particle Schr\"odinger equation (with a real-valued potential $V$ if appropriate),
\begin{equation}\label{Scrodinger equation}
     i \hbar \frac{\partial \Psi(\vx,t)}{\partial t} = -\frac{\hbar^2}{2m} \nabla^2 \Psi(\vx,t) + V\Psi(\vx,t)
\end{equation}
for $\vx\in \Omega$ with the boundary condition 
\begin{equation}\label{ABC}
    \vn(\vx) \cdot \nabla \Psi(\vx,t) = i \kappa \Psi(\vx,t)
\end{equation}
for all $\vx \in\partial\Omega$,
where $\kappa > 0$ is a constant characterizing a property of the detector and $ \vn(\vx)$ is the outward unit normal vector to the surface $\partial\Omega$ at $\vx$. Then 
\be
\PPP\bigl(\vX_D\in d^2\vx, T_D \in dt \bigr)= \vn(\vx) \cdot \vj(\vx,t)\, d^2\vx \, dt
\ee
with the current $\vj$ as in \eqref{jdef}.

The boundary condition \eqref{ABC} ensures that at any boundary point $\vx$, $\vj(\vx,t)$ points outward; as a consequence, a Bohmian trajectory can cross the boundary only outwards. Still, part of the wave function can be reflected back. For example, in the one-dimensional case of a half-line $\Omega=(-\infty,0]$ with boundary $\partial\Omega=\{0\}$, the fraction of the reflected wave for an incident plane wave $\exp(ikx)$ can be quantified by the reflection coefficient
\begin{equation}
   \mathcal{R}_k=\frac{(k-\kappa)^2}{(k+\kappa)^2}. 
\end{equation}
This entails that for the special value of $k=\kappa$ we observe full absorption and zero reflection of the incident wave. As already mentioned in Remark~\ref{rem:hard} in Section~\ref{sec:intro}, for the scattering regime we do not necessarily want to consider \emph{hard} detectors, but want to avoid substantial reflection of the wave function from the detectors; that is why we do not employ the absorbing boundary rule.

To be sure, one would obtain \eqref{cross} from the ABR if one adjusted $\kappa$ so that $\kappa(t)$ equals the $|\vk|$ value of the wave that arrives at time $t$, i.e., $|\vk| = mR/\hbar t$. But that does not correspond to how scattering experiments are actually done.

\subsection{Imaginary Potentials}

Let us explain briefly why imaginary potentials can serve as an effective model of detectors. Imaginary potentials have been used since at least 1940 \cite{Bet40} for modeling absorption of particles, and for modeling detection since at least 1969, when Allcock published a series of three  influential papers~\cite{All69b}. In this model, in the joint wave function of the particle and the detectors, part of the pre-detection wave function will propagate into the region of configuration space where interaction between the particle and some detector can occur. Part of \emph{that} will undergo interaction and, as a consequence, be quickly transported to a region of configuration space with macroscopically different configurations of the detector, and will not return from there (as discussed for hard detectors in \cite{detect-rule}). Thus, the pre-detection part of the wave function will shrink over time as if the particle was absorbed. The wave function $\Psi$ we are considering is the particle factor of the pre-detection part of the full wave function, and thus undergoes a non-unitary evolution; $|\Psi(\vx,t)|^2 \, d^3\vx$ means the probability that at time $t$ the particle has not been detected yet and is located in the volume element $d^3\vx$. An imaginary potential leads to such a non-unitary evolution, in fact one in which $|\Psi|^2$ undergoes a loss at $\vx$ with rate $\tfrac{2}{\hbar}\Im(-V(\vx))|\Psi(\vx,t)|^2$, corresponding to an absorption rate of $\tfrac{2}{\hbar}\Im(-V(\vx))=\tfrac{2}{\hbar}\lambda 1_{\Omega^c}(\vx)$ (where $1_{\Omega^c}$ denotes the characteristic function of the set $\Omega^c$, which is 1 inside $\Omega^c$ and 0 outside).
Correspondingly, the joint probability distribution of the detection observables is given by 
\begin{equation}\label{TDdefIm}
    \mathbb{P}\bigl(\vX_D \in d^3\vx,T_D \in dt\bigr)= \tfrac{2}{\hbar}\lambda 1_{|\vx|>R} \, |\Psi(\vx,t)|^2 d^3\vx \, dt. 
\end{equation} 

\begin{figure}
  \centering
\includegraphics[page=1, width=0.9\textwidth]{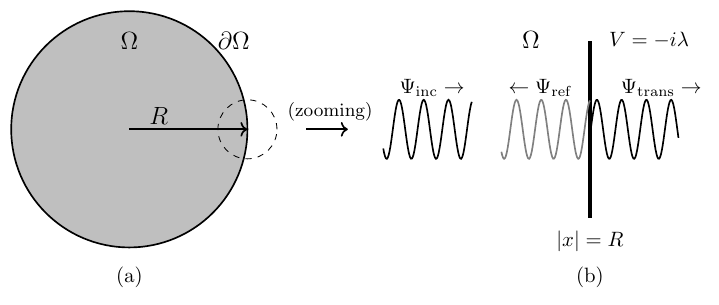}
  \caption{Figure~(a) illustrates a ball of radius \( R \), denoted by \( \Omega \), which is enclosed by detectors positioned along its boundary \( \partial\Omega \). The detectors detect particles that reach this boundary.
Figure~(b) shows a zoomed-in view near the boundary. In the limit \( R \to \infty \), the curved boundary appears g flat, effectively resembling a straight wall.
For the error estimates in the model with imaginary potentials, it plays a role that an incident  plane wave \( \inc \) will, due to the potential step at the boundary \( \partial\Omega \), be partly reflected and partly  transmitted.}
    \label{Set-up}
\end{figure}

\section{Results}\label{Results}

\emph{Setup-} We consider a spinless non-relativistic quantum particle with initial wave function $\Psi_0(\vx)$. We will model detectors surrounding the ball $\Omega =B_{R}$ (see Figure \ref{Set-up}).
Here, $R$ also sets the scale to which $T_D$ and $\vX_D$ should be compared. Since we consider the limit $R\to\infty$ but keep $\Psi_0$ fixed, it follows that $\Psi_0$, when considered on the scale $R$, will be concentrated near the origin.

\subsection{First Model: Imaginary Potential} 
\label{sec:Imresults}

We first model the detectors using an imaginary potential in the region $\Omega^c=\{|\vx|\geq R\}$ outside $S_R$ (where $A^c$ means the complement of any set $A$),
\be\label{Vdef}
V(\vx)=-i\lambda 1_{\Omega^c}(\vx)\,,
\ee
in the 1-particle Schr\"odinger equation for $\Psi=\Psi_{WID}$,
\begin{equation}\label{Schrodinger}
     i \hbar \frac{\partial \Psi(\vx,t)}{\partial t} = -\frac{\hbar^2}{2m} \nabla^2 \Psi(\vx,t) + V\Psi(\vx,t) \,.
\end{equation}

We consider any limit in which $R\to\infty$, $\lambda\to 0$ and $\lambda R \to \infty$. 
The results are equally valid if we first let $R\to\infty$ and then $\lambda\to 0$. 
Our first result is that \eqref{cross} holds in this limit, i.e., that the distribution \eqref{TDdefIm} takes the form \eqref{cross3}. 

Further results concern the Bohmian trajectories, i.e., the solutions $\vQ(t)$ of the equation of motion
\be \label{Bmtraj}
\frac{d\vQ}{dt} = \frac{\vj(\vQ(t),t)}{|\Psi(\vQ(t),t)|^2}
\ee
with $\vj$ as in \eqref{jdef} and random initial position $\vQ(0)$ with distribution
\be \label{inidis}
\PPP\bigl(\vQ(0)\in d^3\vx\bigr) = |\Psi_0(\vx)|^2 \, d^3\vx\,.
\ee
We consider $\vQ(t)$ in two cases, either as $\vQ=\vQ_{WID}$ for $\Psi=\Psi_{WID}$ evolving under the imaginary potential \eqref{Vdef} or as $\vQ=\vQ_{WOD}$ for $\Psi=\Psi_{WOD}$ evolving under the free Schr\"odinger equation. In each case, we define $T_{WID/WOD}$ and $\vX_{WID/WOD}$ as the time and place of first arrival on $S_R$,
\be \label{Twid/wod}
T_{WID} = \inf \bigl\{ t>0:|\vQ_{WID}(t)|\geq R \bigr\}\,,~~ 
\vX_{WID} = \vQ_{WID}(T_{WID})
\ee
and likewise for WOD. We find that with probability approaching 1,
\begin{align}
    \frac{T_{D}-T_{WID}}{R}&=\mathcal{O}(1/\lambda R) \label{TDTWIDerror}\\[3mm]
    \frac{\vX_{D}-\vX_{WID}}{R}&=\mathcal{O}(1/\lambda R), \label{XDXWIDerror}\\[3mm]
    \frac{T_{WID}-T_{WOD}}{R}&=\mathcal{O}(\lambda), \label{TWIDTWODerror}
    \\[3mm]
    \frac{\vX_{WID}-\vX_{WOD}}{R}&= \begin{cases}o(\lambda) & \text{ for spherical surfaces}\\[1mm] \mathcal{O}(\lambda) & \text{ for other shapes as in Remark~\ref{rem:shape}.} \end{cases}.  \label{XWIDXWODerror}
\end{align}
Notably, one can see that the differences in the numerators are large in absolute values but small when compared to the scale of $R$, which justifies the considerations usually made in the scattering regime. It was known before \cite{Daumer,DGMZ06} that the joint distribution of $T_{WOD},\vX_{WOD}$ is asymptotically given by \eqref{cross}, i.e., that \eqref{cross3} holds for $T_{WOD},\vX_{WOD}$ instead of $T_D,\vX_D$. The deviation $(T_D-T_{WOD})/R$ is of the magnitude $\mathcal{O}(1/\lambda R)+\mathcal{O}(\lambda)$ according to \eqref{TDTWIDerror}, \eqref{TWIDTWODerror}; which of the two contributions is larger depends on whether $\lambda$ shrinks faster or slower than $1/\sqrt{R}$.

\begin{rem}\label{rem:TWIDTWODleading}
For $T_{WID}-T_{WOD}$, we can be more explicit: we not only obtain its order of magnitude \eqref{TWIDTWODerror} but can also give the leading order expression:
\begin{align}
    \frac{T_{WID}-T_{WOD}}{R}= \lambda \: \frac{\Im \Bigl[\widehat{\Psi}_0\bigl(\frac{m\vv_0}{\hbar} \bigr) \exp\bigl(i\frac{2mR|\vv_0|}{\hbar}\bigr) \, J \Bigr]}{m|\vv_0|^3 \, \Bigl|\widehat{\Psi}_0\bigl(\frac{m\vv_0}{\hbar}\bigr) \Bigr|^2} + \mathcal{O}(\lambda^2)\,, \label{Tdelayleading}
\end{align}
where $\vv_0=\vv_{WOD}=\vX_{WOD}/T_{WOD}$ is the asymptotic velocity of the trajectory $\vQ_{WOD}$ and 
\begin{equation}\label{Jdef}
    J= \int_0^1 d\rho \ \frac{1}{2-\rho}\exp\left( -i\frac{2mR|\vv_0|}{\hbar \rho} \right) \widehat{\Psi}_0^*\left(\frac{2-\rho}{\rho} \frac{m\vv_0}{\hbar}\right)\,. 
\end{equation}
See after \eqref{Tint} in Section~\ref{Impot}.\hfill$\diamond$\end{rem}

\subsection{Second Model: Zeno Dynamics}
\label{sec:Zenoresults}

\begin{figure}
    \centering
\includegraphics[width=0.5\textwidth]{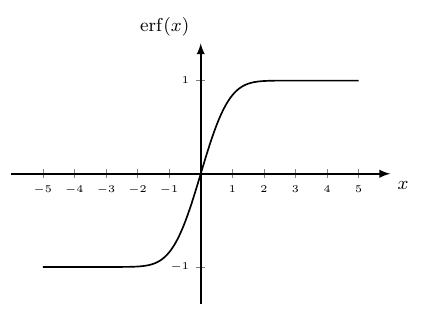}
     \caption{Shape of the error function, $\operatorname{erf}(x)$, as defined in Eq.\ \eqref{erfdef}.}
    \label{erf}
\end{figure}

Secondly, we study the same setup but we model the detectors in a different way. We use repeated nearly-projective measurements of the observable $1_{\Omega^c}$ at non-random times $\timestep, 2\timestep, 3\timestep, \\ \ldots, n\timestep, \ldots$. The reason we use a \emph{nearly-projective} measurement rather than an \emph{exactly projective} one is to make smaller (though still non-zero) the part of the collapsed wave function traveling backwards (analogous to reflection in the imaginary potential case). More precisely, ``nearly-projective'' means that the quantum measurement is only approximately but not exactly one of the PVM (projection-valued measure) consisting of the multiplication operator $1_{\Omega}$ (for ``no detection'') and the complementary projection $1_{\Omega^c}$ (for ``detection''); rather, it is a quantum measurement of the POVM (positive-operator-valued measure) consisting of the two multiplication operators
\begin{align}
    &\hat{P}_\sigma=
        \frac{1}{4}\left( 1- \operatorname{erf}\Bigl(\frac{|\vx|-R}{2\sigma}\Bigr)\right)^2~~~~\text{(for ``no detection'')} \label{Pdef}\\
    &\hat{I}-\hat{P}_\sigma
    ~~~~\text{(for ``detection''),} \label{IPdef}
\end{align}
where ``erf'' denotes the error function
\be \label{erfdef}
\operatorname{erf}(x) =  2 \int_0^x \frac{1}{\sqrt{\pi}}e^{-y^2} \, dy\,,
\ee
the anti-derivative of the Gaussian density with width $1/\sqrt{2}$, normalized so that it approaches $\pm 1$ as $x\to\pm\infty$, see Figure~\ref{erf}. 

\begin{figure}
    \centering
    \includegraphics[width=10cm, height= 20cm]{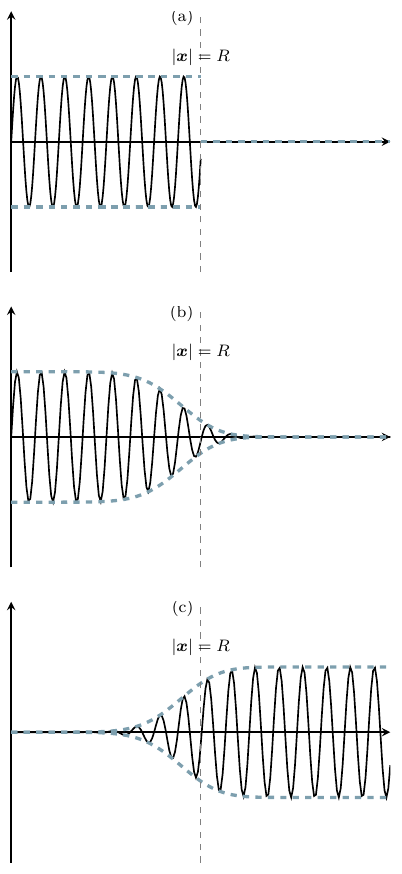}
     \caption{Illustration of the nearly-projective measurement used in the Zeno dynamics at time $n\timestep$. Shown are $\pm|\Psi|$ (dashed) and $\Re \, \Psi$ (black) as a function of $|\vx|$ in a neighborhood of $\vx_0$ with $|\vx_0|=R$ in several cases with $\Psi_{WOD}=\exp(i\vk\cdot \vx)$: (a) for $\Psi= 1_\Omega \Psi_{WOD}$ i.e., after a projective measurement (no detection); (b) $\Psi=\hat{P}_\sigma \Psi_{WOD}$ as in \eqref{Pdef}, i.e., after a nearly-projective measurement (no detection); (c) $\Psi=(\hat{I}-\hat{P}_\sigma)\Psi_{WOD}$ as in \eqref{IPdef}, i.e., after a nearly-projective measurement (detection).
    }
    \label{softstep}
\end{figure}

Thus, $\hat{P}_\sigma$ is the multiplication by a function that is a smooth approximation to the step function $1_{\Omega}$, a ``soft step'' with width $\sigma$ as shown in Figure~\ref{softstep}. 

\begin{rem}
The error function used here belongs to a broader class of functions known as squash functions---functions typically used to compress or ``squash'' large input domains into a bounded interval. Formally, a squash function is defined as an increasing, continuous function 
\be
f: \mathbb{R} \to (a,b) \quad \text{such that} \quad \lim_{x \to -\infty} f(x) = a, \quad \lim_{x \to \infty} f(x) = b.
\ee
Common examples, apart from error function, include $1/(1+e^{-x}) \in (0,1)$, the arc tangent $\arctan(x)\in (-\frac{\pi}{2},\frac{\pi}{2})$, and the hyperbolic tangent $\tanh(x) \in (-1,1)$. Among these, the error function is particularly well-suited for our purposes for two reasons: first, like a Gaussian, it has light tails, i.e., it decays quickly as $x\to\pm\infty$; second, it is easy to compute the time evolution under the free Schr\"odinger equation of $\operatorname{erf}(x)\, e^{ikx}$, using the fact that the derivative of erf is a Gaussian whose Schr\"odinger evolution is well known; see Section~\ref{Zeno}.
\hfill$\diamond$\end{rem}

We turn to the collapse of $\Psi$. For any function $f(t)$, we use the notation
\be
f(t\pm) := \lim_{\varepsilon \to 0} f(t\pm |\varepsilon|)
\ee
for the limiting value right after (before) $t$.
For the measurement at time $n\timestep$, we choose the width
\be\label{sigmandef}
\sigma:=\sigma_n=n \sigma_1 \text{  with } \mathcal{O}\Bigl(\frac{\timestep}{R}\Bigr)\ll \sigma_1 \ll \mathcal{O}\Bigl(\frac{\timestep^2}{R}\Bigr)\,.
\ee 
The probability of detection is thus $\langle \Psi_{n\timestep-}|\hat{I}- \hat{P}_{\sigma_n}| \Psi_{n\timestep-} \rangle$.
After each of these measurements, in case of no detection, the post-measurement wave function gets collapsed to 
\begin{equation}\label{collapsedpsi0}
    \Psi_{n\timestep+}= \frac{\hat{P}^{1/2}_{\sigma_n}\Psi_{n\timestep-}}{\|\hat{P}^{1/2}_{\sigma_n}\Psi_{n\timestep-}\|}.
\end{equation}
In the event of a detection at $n\timestep$, we set $T_D:=n\timestep$, apply an ideal position measurement immediately afterwards, and call the outcome $\vX_D$, so
\be \label{PDdefzeno}
\PPP(\vX_D \in d^3\vx|T_D=n\timestep) = |\Psi_{n\timestep+}(\vx)|^2 \, d^3\vx
\ee
with 
\be \label{collapsedpsi1}
\Psi_{n\timestep+}= \frac{(\hat{I}-\hat{P}_{\sigma_n})^{1/2}\Psi_{n\timestep-}}{\|(\hat{I}-\hat{P}_{\sigma_n})^{1/2}\Psi_{n\timestep-}\|}.
\ee

Our results apply to any limit in which $R\to\infty$, $\timestep \rightarrow \infty$, $\timestep/R\rightarrow 0$, $R\sigma_1/\timestep\to\infty$, and $R\sigma_1/\timestep^2\to 0$. The first result is that \eqref{cross3} holds again.

The second result is concerned with the magnitude of the deviations $T_D-T_{WID}$ etc.; since these quantities involve the Bohmian trajectories $\vQ(t)$, defined in \eqref{Bmtraj} and \eqref{inidis}, we need to specify how the nearly-projective measurements act for a given Bohmian position $\vQ(t)$. Here again, we use the notation $\vQ(t)=\vQ_{WOD}(t)$ for $\Psi=\Psi_{WOD}$ evolving under the free Schr\"odinger equation,  and $\vQ(t)=\vQ_{WID}(t)$ for $\Psi=\Psi_{WID}$ evolving in accordance with the repeated nearly-projective measurements of the observable $1_{\Omega}$. x
We model the latter as follows, similarly to \cite{Tum11}. At each time $n\timestep$ of attempted detection, a real-valued random variable $Z$ is chosen independently of the past (including independently of $\vQ_{WID}(n\timestep-)$) with probability density
\be\label{Zdisbn}
p_{\sigma_n}(z)= \frac{1}{2\sqrt{\pi}\sigma_n}\left( 1+ \operatorname{erf}\left( \frac{z}{2\sigma_n}\right)\right)\exp\left(- \frac{z^2}{4\sigma^2_n}\right) \,.
\ee
The quantity $Z$ serves for representing the fuzziness of the measurement. We define that the outcome of the experiment is 
\begin{align} 
\text{``no detection'' if} \quad &|\vQ_{WID}(n\timestep-)| + Z < R \,, \\
\text{``detection'' if} \quad  
&|\vQ_{WID}(n\timestep-)| + Z  \geq R \,.
\end{align}
We show in Section~\ref{sec:2ndZeno} that this definition leads to exactly the probability $\langle \Psi_{n\timestep-}|\hat{P}_{\sigma_n}|\Psi_{n\timestep-}\rangle$ for ``no detection,'' as it should.
 
The Bohmian position is left unchanged during the collapse, and from time $n\timestep+$ onwards the Bohmian position moves according to \eqref{Bmtraj}, with $\Psi=\Psi_{n\timestep+}$ given by \eqref{collapsedpsi1} or \eqref{collapsedpsi0}. We show in Section~\ref{sec:2ndZeno} that, conditioned on either ``no detection'' or ``detection'' (whichever applies), $\vQ_{WID}(n\timestep+)=\vQ_{WID}(n\timestep-)$ has distribution density $|\Psi_{n\timestep+}|^2$. This completes the definition of the model of detection.\footnote{At time $n\timestep$, the Bohmian velocity will not have a discontinuity either, as $\Psi_{n\timestep-}$ gets multiplied by a real valued function \eqref{Pdef} or \eqref{IPdef}, leading to no additional contribution to the total phase. While discontinuities may occur in higher-order time derivatives of $\vQ(t)$, these do not affect the deviations in the Bohmian variables beyond the errors listed in \eqref{TDTWIDerrorZeno}--\eqref{XWIDXWODerrorZeno}.}  

We find that the deviations associated with the Bohmian arrival variables are, with probability approaching 1,
\begin{align}
    \frac{T_{D}-T_{WID}}{R}&=\mathcal{O}(\timestep/R) \label{TDTWIDerrorZeno}\\[3mm]
    \frac{X_{D}-X_{WID}}{R}&=\mathcal{O}(\timestep/R), \label{XDXWIDerrorZeno}\\[3mm]
    \frac{T_{WID}-T_{WOD}}{R}&=o(\timestep/R), \label{TWIDTWODerrorZeno}
    \\[3mm] 
    \frac{X_{WID}-X_{WOD}}{R}&= o(\timestep/R).
    \label{XWIDXWODerrorZeno}
\end{align}
Here again, we see that when compared to $R$, the deviations tend to 0. Note here though, if we had considered $\timestep \rightarrow 0$, then we would have seen no detection, $T_D \rightarrow \infty$ (quantum Zeno effect). 

\begin{rem} (Energy-time uncertainty relation) Are our results in agreement with an energy-time uncertainty relation of the form $\sigma_E \, \sigma_{T_D}\geq \hbar/2$? Yes, because such a relation was proved in \cite{KRSW12} for $T_D$ the time of absorption in an imaginary potential. On the other hand, we have seen that the distribution of $T_D/R$ is asymptotically given by that of the quantum observable $m|\hat{\vp}|^{-1}= \sqrt{m/2} \hbar \hat{H}^{-1/2}$, which commutes with the Hamiltonian $\hat{H}=-(\hbar^2/2m) \nabla^2$. This raises the question, since commuting observables are not subject to an uncertainty relation, how is this commutation compatible with the general energy-time uncertainty mentioned before? It is compatible for two reasons: first, because $T_D$ differs from the observable $Rm|\hat{\vp}|^{-1}$ by $\mathcal{O}(1)$, and second, because the scaling factor $R$ needs to be taken into account: the uncertainty $\sigma_{T_D/R}$ of $T_D/R$ is $R^{-1} \sigma_{T_D}$, so $\sigma_E \, \sigma_{T_D/R}\geq \hbar/2R$, which vanishes in the limit $R\to\infty$ that we are considering.
\hfill$\diamond$\end{rem}

\section{Derivation in the Case of Imaginary Potential}\label{Impot}

We now turn to the justification of our claims.
In this section, we model the presence of detectors in the region $\Omega^c$ using an imaginary potential of the form $V(\vx)=-i\lambda 1_{\Omega^c}(\vx)$. As mentioned before, we consider any limit in which $R\rightarrow \infty$, $\lambda \rightarrow 0$, and $\lambda R \rightarrow\infty$. We remark that our results are also valid if we let $R\rightarrow \infty$ first and then take $\lambda \rightarrow0$.

The basic strategy of our analysis is to ``zoom in'' to a neighborhood of a point on the detecting surface as in Figure~\ref{Set-up}. Eq.~\eqref{asymppsi} tells us that, in this neighborhood, the incident wave is essentially a plane wave. Due to the largeness of $R$, the local curvature of the surface vanishes, so we can take the surface to be a plane, and $\Omega$ a half space; the Schr\"odinger equation becomes that of a ``single step'' potential (here imaginary). Since the $\vk$ vector points in the radial direction, which is perpendicular to the surface, the radial direction decouples from the transversal directions, and the problem becomes essentially 1-dimensional. (As an alternative strategy, one could determine the eigenfunctions of the 3D Hamiltonian with the spherically symmetric potential $V$ for fixed $R$ and $\lambda$, expand $\Psi$ in those, and determine the asymptotics of its time evolution from there. We plan to give more details about this approach in a future work, to avoid the technicalities of the derivation here.)

We will show that,
upon encountering the potential at the boundary, a fraction of the wave is reflected back into $\Omega$, while the remainder is transmitted into the detector region $\Omega^c$. We will see that the reflected wave, which is an inward-propagating disturbance, leads to a \emph{delay} in the arrival of the particle at the boundary $\partial\Omega$, as compared to the case without detectors. This delay accounts for the deviation $T_{{WID}} - T_{{WOD}}$. Meanwhile, the transmitted portion continues into the detector region and governs the dynamics leading to detection at a random position $X_D$ and time $T_D$ within $\Omega^c$.

\bigskip

{\it Computation of reflection and transmission coefficients-}
In order to compute the amplitude of the reflected wave from an imaginary potential step of height $\lambda$, we consider the 1d stationary solutions to the Schr\"odinger equation with $V(x)=-i \lambda 1_{x>0}$ (putting the origin at the step). Since the general solution of the eigenfunction equation of the Laplacian plus a constant potential is $A\exp(ikx)+B\exp(-ikx)$, the desired stationary solutions must be of the form
\be\label{stationarystep}
\psi(x) = \begin{cases}
\psi_{\leq}(x) := A e^{ikx} + B e^{-ikx} & \text{for }x\leq 0\\
\psi_{\geq}(x):= C e^{iKx} + D e^{-iKx} & \text{for }x\geq 0
\end{cases}
\ee
with $\hbar^2 k^2/2m = \hbar^2 K^2/2m - i \lambda$ (to ensure that $\psi_{\leq}$ and $\psi_{\geq}$ are compatible with the same eigenvalue). 
Since $k$ is real, it follows that $K$ is complex, and
\be
 K=k + i \frac{m}{\hbar^2 k}\lambda + \mathcal{O}(\lambda^2)\,. 
\ee
The matching conditions
\be\label{matching}
\psi_{\leq}(x=0)=\psi_{\geq}(x=0)\,,~~~\partial_x\psi_{\leq}(x=0) = \partial_x \psi_{\geq}(x=0)
\ee
lead to relations between the coefficients $A,B,C,D$. We are interested in the solutions with $D=0$ (an incoming wave from the left) and can take $A=1$ without loss of generality. The matching conditions, $1+B=C$ and $ik(1-B)=iKC$, then lead to
\begin{align}
B_k:=B&=\frac{k-K}{k+K}=-i\frac{m\lambda}{2\hbar^2 k^2}+\mathcal{O}(\lambda^2)\\
C_k:=C&=\frac{2k}{k+K}=1- i \frac{m\lambda}{2\hbar^2 k^2}+\mathcal{O}(\lambda^2)\,.
\end{align}
These coefficients $B_k$ and $C_k$ can now be applied to the radial direction of the 3D problem. 

This reasoning remains valid for arbitrary, non-spherical surfaces (see Section~\ref{arbsurf}); however, in those cases, the orientation of the boundary must be explicitly accounted for.

\bigskip

{\it Asymptotic form of $\Psi_{WID}$-}
Analogously to Eq.\ \eqref{asymppsi}, the asymptotic form of the wave function in the presence of the imaginary potential at large $t$ is given by
\be
\Psi_{WID}(\vx,t) = \begin{cases}
    \inc(\vx,t) + \refl(\vx,t) & \text{if }|\vx|<R\\
    \trans(\vx,t) &\text{if }|\vx|\geq R\,,
\end{cases}
\ee
where the incident wave $\inc$ is given by \eqref{asymppsi}, and the reflected and transmitted waves are
\begin{align}
   \refl(\vx,t)&\approx \left(\frac{m}{i\hbar t}\right)^{3/2}B_{|\vk'|}  \,\frac{2R-|\vx|}{|\vx|} \,\widehat{\Psi}_0 (\vk^\prime) \, 
    \exp\bigl(i2|\vk'| R-i\vk'\cdot \vx-i\omega' t\bigr)
    \label{PsiR}\\
    \trans(\vx,t)&\approx \left(\frac{m}{i\hbar t}\right)^{3/2}C_{|\vk|} \, \widehat{\Psi}_0(\vk) \, 
    \exp \biggl(-\frac{\lambda t (|\vx|-R)}{\hbar |\vx|}\biggr) \, \exp(i\vk\cdot\vx -i\omega t) \label{PsiT}
\end{align}
with
\be
\vk^\prime= \frac{m}{\hbar} \frac{2R-|\vx|}{t} \frac{\vx}{|\vx|},~~\omega'=\frac{\hbar\vk^{\prime 2}}{2m},
\ee
and $\vk,\omega$ as in \eqref{asymppsi}. 

Formula \eqref{PsiR} can be understood as follows: it refers to the Fourier mode that arrives at $\vx$ at time $t$ after having reached $R$ and gone back to $|\vx|$. Its original wave vector is $\vk'$ and amplitude $\widehat{\Psi}_0(\vk')$, but after reflection it propagates inwards, which leads to the minus sign in $-i\vk' \cdot \vx$. The phase accumulated during the travel to $R$ and back to $|\vx|$ leads to the additional phase $2|\vk'|R$; the amplitude and phase of the wave  change due to the reflection at $R$ by the factor $B_{|\vk'|}$. The factor $(2R-|\vx|)/|\vx|$ has geometric reasons: since the reflected wave is moving radially inward, the amount of probability it would carry, $|\widehat{\Psi}_0(\vk')|^2 d^3\vk'$, will be concentrated on smaller and smaller $\vx$-volumes $d^3\vx$.

Formula \eqref{PsiT} can be understood as follows: when crossing $S_R$, the wave changes by a factor $C_{|\vk|}$, continues as a plane wave with vector $\vk$, but is then subject to an exponential decay $\exp(-\lambda \Delta t/\hbar)$ at rate $\lambda/\hbar$. For the part of $\Psi$ found at $|\vx|>R$ at time $t$, since it is still propagating outward at the same speed, the time it crossed the sphere must have been $t_{\text{cross}}=tR/|\vx|$, so $\Delta t=t-t_{\text{cross}}= t(|\vx|-R)/|\vx|$.

Note that alternatively, formulas \eqref{PsiR} and \eqref{PsiT} could be obtained through a stationary phase calculation. This approach makes use of the eigenfunctions of the 3D Hamiltonian with the spherically symmetric potential $V$ for fixed $R$ and $\lambda$, mentioned before. We leave this approach for future work. 

\bigskip

{\it Detection time and place-}
The time $T_{D}$ and place $\vX_{D}$ at which the particle will be detected are random. Once the particle enters the region $\Omega^c$, the detection event occurs at a constant rate $2\lambda/\hbar$, so that $T_D-T_{WID}$ is exponentially distributed with expectation $\hbar/2\lambda$. In particular, 
\be\label{Td-TwidIM}
    T_D-T_{WID}= \mathcal{O}(1/\lambda)\,.
\ee
Since velocities are of order 1, also
\be
    \vX_D-\vX_{WID}= \mathcal{O}(1/\lambda).
\ee
$T_{WID}-T_{WOD}$ will be studied after \eqref{Tint} below.

We now compute the limiting joint distribution of $T_D$ and $\vX_D$
from \eqref{TDdefIm}: Let $\lim$ denote a limit in which $\lambda\to0$, $R\to\infty$, and $\lambda R \to \infty$. Then
\begin{align}\label{PDImpot}
&\lim\mathbb{P}\biggl(\frac{\vX_{D}}{|\vX_D|}\in d^2\vu,~ \frac{|\vX_D|}{R} \in d\rho,~  \frac{T_{D}}{R}\in d\timescaled\biggr) = \nonumber\\
   &\stackrel{\eqref{TDdefIm}}{=} \lim \frac{2\lambda}{\hbar} 1_{\rho\geq 1} \Bigl|\trans(R\rho \vu, R\timescaled)\Bigr|^2 R^4\rho^2 \, d\rho \, d^2\vu\, d\timescaled\\
    &\stackrel{\eqref{PsiT}}{=} \lim \frac{2\lambda}{\hbar} 1_{\rho\geq 1} \Bigl( \frac{m}{\hbar R\timescaled}\Bigr)^3 \: |C_{|\vk|}|^2 \: \Bigl|\widehat{\Psi}_0 \Bigl(\frac{m\rho}{\hbar \timescaled}\vu \Bigr) \Bigr|^2 \exp \left(-\frac{2\lambda R\timescaled(R\rho-R)}{\hbar R\rho}\right) \, R^4\rho^2 \, d\rho \, d^2\vu\, d\timescaled\\
    &\:=\lim \frac{m^3\rho^2}{\hbar^3\timescaled^4}\Bigl|\widehat{\Psi}_0\Bigl(\frac{m \rho}{\hbar \timescaled} \vu\Bigr) \Bigr|^2  \biggl[ 1_{\rho\geq 1}\frac{2\lambda R\timescaled}{\hbar} \exp\Bigl(-\frac{2\lambda R\timescaled}{\hbar} \frac{\rho-1}{\rho}\Bigr) \biggr] d^2\vu \, d\rho \, d\timescaled\\
    &\:= \frac{m^3}{\hbar^3 \timescaled^4} \,  \Bigl|\widehat{\Psi}_0\Bigl(\frac{m\vu}{\hbar \timescaled}\Bigr)\Bigr|^2 \, \delta(\rho-1)\, d^2\vu \, d\rho\, d\timescaled\,,
\end{align}
where we used that $|C_{|\vk|}|^2= 1+ \frac{\lambda^2}{4|\vk|^4}\to 1$, that
\be\label{expdelta}
\lim_{\alpha \to \infty} \Bigl( 1_{y\geq 0} \: \alpha \: e^{-\alpha y} \Bigr) = \delta(y)
\ee
(applied to $\alpha=2\lambda R\timescaled/\hbar$ and $y=(\rho-1)/\rho$), that $\delta(1-1/\rho)=\delta(\rho-1)$, and that $\rho^2 \delta(\rho-1)=\delta(\rho-1)$. Indeed, \eqref{expdelta} can be derived by noting that for any smooth and compactly supported test function $f$,
\be
\lim_{\alpha \to \infty} \int_{-\infty}^\infty dy \, 1_{y\geq 0} \, \alpha \, e^{-\alpha y} \, f(y) = f(0)\,,
\ee
the defining property of the delta function.
This completes the derivation of \eqref{cross3}.

\bigskip

{\it Influence on Bohmian trajectories-} Within the sphere $S_R$, the Bohmian velocity field can be expressed as
\begin{align}
    \vv_{WID}&= \frac{\hbar}{m} \frac{\Im[(\inc + \refl)^*\nabla(\inc + \refl)]}{|\inc + \refl|^2}\\
    &\approx \vv_{WOD} + \frac{\hbar}{m} \left[\frac{\Im[\refl^*\nabla\inc + \inc^*\nabla\refl]}{|\inc|^2} - \frac{\vv_{WOD}}{|\inc|^2}2\Re(\refl^*\inc) \right],
\end{align}
where
\be
\vv_{WOD} = \frac{\hbar}{m} \frac{\Im[\inc^* \nabla \inc]}{|\inc|^2}
\ee
represents the Bohmian velocity in the absence of detectors. Since $\refl$ is of order $\lambda$, terms involving $\refl^* \nabla \refl$ or $\refl^* \refl$ are of order $\lambda^2$ and therefore negligible. We retain only the first-order corrections in $\lambda$, i.e., terms of $\mathcal{O}(\lambda)$, while higher-order contributions are approximated as zero. Note that
\be\label{vdelaydef}
\vv_{\mathrm{delay}}:=\vv_{WOD}-\vv_{WID}= \mathcal{O}(\lambda)\,.
\ee
It turns out that $\vv_{\mathrm{delay}}\geq 0$, so the particle slows down due to $\refl$ (as if it felt headwinds).

It is known \cite{RDM05} that free Bohmian particles, guided by $\Psi_{WOD}$, asymptotically move along straight lines; this can be understood by noting that a plane wave with wave vector $\vk$ leads to straight Bohmian trajectories with velocity $\hbar\vk/m$, which furthermore is exactly the velocity at which the Fourier mode $\vk$ propagates, so that the Bohmian particle follows the Fourier mode and continues with velocity $\hbar \vk/m$. In the present case, although $\Psi_{WID}$ does not evolve freely, $\vv_{WID}(\vx)$ for $|\vx|<R$ is still pointing in the radial direction up to $o(\lambda)$ corrections, with the consequence that the Bohmian particle moves radially up to $o(\lambda)$. In addition, the velocity is given by $\vv_{WOD}$ up to the $\mathcal{O}(\lambda)$ correction $\vv_{\mathrm{delay}}$, which allows us to compute the total time delay $T_{WID}-T_{WOD}$ and thereby confirm Remark~\ref{rem:TWIDTWODleading} as follows: Let $\vu$ be the unit vector in the outward radial direction, so $\vu\cdot \vv_{WOD}= |\vv_{WOD}|$. Note that the time $T$ needed for traveling the distance $R$ with speed $v(x)$ is
\be\label{Tint}
T=\int_0^R \frac{dx}{v(x)}\,,
\ee
and that
\be
\frac{1}{\vu\cdot \vv_{WID}}-\frac{1}{\vu\cdot \vv_{WOD}} = \frac{\vu\cdot \vv_\mathrm{delay}}{(\vu\cdot \vv_{WOD})^2} + \mathcal{O}(\lambda^2)\,.
\ee
Let $\vv_0:=\vv_{WOD}$ be the constant velocity of the trajectory without detectors, so $\vQ_{WOD}(t)=\vv_0 t$ is the trajectory to zeroth order in $\lambda$. Then
\begin{align}\label{Twid-TwodIM}
    T_{WID}-T_{WOD}
    &\approx \frac{1}{|\vv_0|^2}\int_0^R d r \, \vu\cdot \vv_{\mathrm{delay}}(r\vu,r/|\vv_0|) \\
    &= \frac{1}{|\vv_0|^2}\int_0^R dr\, \biggl( \frac{\hbar}{m} \frac{\Im[\refl^*\partial_r \inc + \inc^*\partial_r \refl]}{|\inc|^2} - \frac{|\vv_0|}{|\inc|^2}2\Re(\refl^*\inc ) \biggr)\,,
\end{align}
where $\partial_r=(\vx/|\vx|)\cdot\nabla$ means the radial derivative. We find that, for $\vx=\mathcal{O}(R)$ and $t=\mathcal{O}(R)$,
\begin{align}
\partial_r \inc(\vx,t)
&= \Bigl[ i\frac{m|\vx|}{\hbar t}+ \mathcal{O}(1/R) \Bigr]\inc\\
\partial_r \refl(\vx,t)
&= \Bigl[-i\frac{m(2R-|\vx|)}{\hbar t}+ \mathcal{O}(1/R) \Bigr]\refl \,.
\end{align}
Inserting this, writing
$\vv_0:=\vv_{WOD}$ and noting that $t\approx r/|\vv_0|$, we find that, with $J$ given by \eqref{Jdef},
\be
    T_{WID}-T_{WOD} \approx \frac{\lambda R}{m|\vv_0|^3 \, |\widehat{\Psi}_0(m\vv_0/\hbar)|^2}\Im \biggl[\widehat{\Psi}_0\Bigl(\frac{m\vv_0}{\hbar}\Bigr) \exp\Bigl(i\frac{2mR|\vv_0|}{\hbar}\Bigr) J \biggr] \,,
\ee
This proves \eqref{Tdelayleading}.
Notably, the presence of detectors affects the arrival time considerably: since $J=\mathcal{O}(1)$, $T_{WID}-T_{WOD}=\mathcal{O}(\lambda R)$. Deviations in the arrival places, in contrast, are $o(\lambda R)$: indeed, they arise from subleading corrections to $\Psi$ in \eqref{asymppsi} and \eqref{PsiR} (small compared to $\lambda$), which can slightly change the direction $\vQ(t)/|\vQ(t)|$ at any point during the (essentially radial) flight from the origin to $S_R$; this change can get amplified by a factor $R$ during the remaining flight.

\begin{rem}
Alternatively, we can also directly provide explicit expressions for the joint probability distributions associated with the pairs of random variables $(T_{WID}, \vX_{WID})$ as well as $(T_{WOD}, \vX_{WOD})$: Indeed, it is well known that if $\vj$ is everywhere outward-pointing, then $\vn\cdot\vj$ is the probability density of arrival. Thus,
\begin{align}
&\lim\mathbb{P}\biggl(\frac{\vX_{WOD}}{|\vX_{WOD}|}\in d^2\vu,~ \frac{|\vX_{WOD}|}{R} \in d\rho,~  \frac{T_{WOD}}{R}\in d\timescaled\biggr) = \nonumber\\
    &\qquad= \delta(\rho-1) \, d\rho \, \lim \vu \cdot \vj (R\vu,R\timescaled) \: R^3 d^2\vu \, d\timescaled\\
    &\qquad= \delta(\rho-1)\, d\rho\, \lim \tfrac{\hbar}{m}\Im \left[ \inc^* \vu\cdot\nabla \inc\right] \Big|_{(R\vu,R\timescaled)} \, R^3 d^2\vu \, d\timescaled \\
    &\qquad \stackrel{\eqref{asymppsi}}{=} \frac{ m^3}{\hbar^3 \timescaled^4} \Bigl|\widehat{\Psi}_0\Bigl(\frac{m\vu}{\hbar \timescaled}\Bigr)\Bigr|^2 \, \delta(\rho-1)\,d ^2\vu \, d\rho \, d\timescaled,\label{Pwod}\\[2mm]
    &\lim\mathbb{P}\biggl(\frac{\vX_{WID}}{|\vX_{WID}|}\in d^2\vu,~ \frac{|\vX_{WID}|}{R} \in d\rho,~  \frac{T_{WID}}{R}\in d\timescaled\biggr) = \nonumber\\
    &\qquad = \delta(\rho-1) \, d\rho\,  \lim \tfrac{\hbar}{m}\Im \Bigl[ (\inc + \refl)^* \: \vu\cdot \nabla (\inc + \refl)\Bigr] \bigg|_{(R\vu, R\timescaled)} R^3\,d^2\vu\, d\timescaled \label{Pwid}  \\
    &\qquad = \eqref{Pwod}\,. \nonumber
\end{align}
\hfill$\diamond$
\end{rem}
This concludes the analysis of the setup involving imaginary potentials. We now turn to the other detector model and derive analogous results within that framework.

\section{Derivation in the Case of Zeno Dynamics}\label{Zeno}

The approach here involves repeated nearly-projective measurements performed at times
$\timestep, 2\timestep, 3\timestep,\\ \ldots, n\timestep,
\ldots$, where each time the observable (approximately) $1_{\Omega^c}$ is measured. 

\subsection{First Approximation}
\label{sec:firstapprox}

As a \emph{first approximation}, we neglect any boundary effects that arise from the step (sharp or soft) in Figure~\ref{softstep}: then the contribution to $\Psi$ from the Fourier mode with wave vector $\vk$ will be located near $\vx=\hbar \vk t/m$ at time $t$, and a projective measurement given by multiplication with $1_{\Omega^c}$ will remove some of these contributions but not disturb the others. In this approximation, therefore,
\be
\Psi(\vx,n\timestep+\Delta t)=
\begin{cases}
\mathcal{N}_n\left(\frac{m}{i\hbar t}\right)^{3/2} \widehat{\Psi}_0(\vk) \, \exp(i\vk\cdot \vx) \, \exp\left(-i\omega t\right)
&\text{if }|\vx|<R+R\Delta t/n\timestep\\
0 &\text{otherwise}
\end{cases}
\ee
for $0\leq \Delta t<\timestep$ with (as usual) $\vk=m\vx/\hbar t$, $\omega=\hbar \vk^2/2m$, $t=n\timestep+\Delta t$, and normalizing factor
\be\label{calNn}
\mathcal{N}_n=\Biggl( \int d^3\vk \: 1_{\hbar |\vk|n\timestep\leq mR} \: |\widehat{\Psi}_0(\vk)|^2 \Biggr)^{-1/2} \,.
\ee
This leads to \eqref{cross3} using the limit $\timestep \rightarrow \infty$, $R \rightarrow \infty$, $\frac{\timestep}{R} \rightarrow 0$, as well as to \eqref{TDTWIDerrorZeno} and \eqref{XDXWIDerrorZeno}. For $T_{WID}-T_{WOD}$, this approximation yields the estimate 0.

However, it is not clear that boundary effects arising from the spread of the step can be ignored; in fact, between the time $n\timestep$ and $(n+1)\timestep$ (if the outcome at the $n^{\text{th}}$ measurement is zero), the post-measurement state $\Psi_{n\timestep+}$ has a considerable amount of time to spread. During this time, also the step could spread, inward as well as outward, and create problems. To clearly justify our considerations we study a refined, second approximation.

\subsection{Second Approximation}
\label{sec:2ndZeno}

\emph{Type of approximation-} 
We described this approximation in Section~\ref{sec:Zenoresults}: it consists of approximating the exact (sharp) projective measurement of the observable $1_{\Omega^c}$ by a generalized quantum measurement involving a smooth function (i.e., we replace the sharp step with a soft transition modeled by the error function as defined in \eqref{erfdef}). At time $n\timestep$, based on the measurement outcome ``detection'' or ``no detection,'' the wave function collapses according to \eqref{collapsedpsi1} or \eqref{collapsedpsi0}, respectively. In particular, at time $n\timestep$ in case of ``no detection,'' the post-measurement wave function is 
\begin{equation}
    \Psi(\vx,n\timestep+) = \mathcal{N}\frac{1}{2} \left( 1 - \operatorname{erf} \left( \frac{|\vx| - R}{2 \sigma_n} \right) \right)\Psi(\vx,n\timestep-)
\end{equation}
with $\mathcal{N}$ a normalizing factor. Now consider a neighborhood of size $\ll R$ around some $\vx_0\in S_R$; if $\Psi(\vx,n\timestep-)$ can there be approximated, as $\Psi_{WOD}$ can according to \eqref{asymppsi}, by a plane wave $c e^{i\vk_0\cdot\vx}$ with $\vk_0=m\vx_0/\hbar t$ (and we will see soon that it can), then
\be\label{Psierfplane}
\Psi(\vx,n\timestep+) \approx \frac{\mathcal{N}c}{2} \left( 1 - \operatorname{erf} \left( \frac{|\vx| - R}{2 \sigma_n} \right) \right)e^{i\vk_0\cdot \vx}\,.
\ee

\bigskip

\emph{Time evolution-} To compute the time evolution of this function in this neighborhood up to time $(n+1)\timestep$, we need the time evolution of the error function, and for this purpose begin with that of the Gaussian:
\be\label{evolvedGaussian}
e^{-i\hat{H}\Delta t/\hbar}\biggl[ \exp\Bigl(-\frac{x^2}{4\sigma^2}\Bigr) \biggr] = 
\frac{\sigma}{ \sigma(\Delta t)}\exp\Bigl(-\frac{x^2}{4\sigma(\Delta t)^2}\Bigr)
\ee
with Hamiltonian $\hat{H}=-(\hbar^2/2m)\partial_x^2$ and complex
\be\label{sigmaDeltatdef}
\sigma(\Delta t) = \sqrt{\sigma^2+\frac{i\hbar \Delta t}{2m}}\,.
\ee
Since $\partial_x$ commutes with $\hat{H}$,
\begin{align}
\partial_x e^{-i\hat{H}\Delta t/\hbar}\Bigl[ \operatorname{erf}\Bigl(\frac{x}{2\sigma}\Bigr) \Bigr] 
&= e^{-i\hat{H}\Delta t/\hbar}\partial_x \Bigl[ \operatorname{erf}\Bigl(\frac{x}{2\sigma}\Bigr) \Bigr] \\
&= \frac{1}{\sqrt{\pi}\sigma} e^{-i\hat{H}\Delta t/\hbar} \Bigl[ \exp\Bigl(-\frac{x^2}{4\sigma^2}\Bigr) \Bigr] \\
&\! \stackrel{\eqref{evolvedGaussian}}{=} \frac{1}{\sqrt{\pi} \sigma(\Delta t)}\exp\Bigl(-\frac{x^2}{4\sigma(\Delta t)^2}\Bigr) \\
&= \partial_x\operatorname{erf}\Bigl(\frac{x}{2\sigma(\Delta t)}\Bigr)\,,
\end{align}
so
\be\label{evolvederf}
e^{-i\hat{H}\Delta t/\hbar}\Bigl[ \operatorname{erf}\Bigl(\frac{x}{2\sigma}\Bigr) \Bigr] = \operatorname{erf}\Bigl(\frac{x}{2\sigma(\Delta t)}\Bigr)+C\,,
\ee
where $C$ is some complex constant and the \emph{complex} error function is defined as the anti-derivative of the appropriate complex Gaussian, or equivalently as the analytic continuation of the real error function. 
The complex error function is an entire function with the properties $\operatorname{erf}(-z)=-\operatorname{erf}(z)$ and, for any $-\tfrac{\pi}{4}<\theta<\tfrac{\pi}{4}$,
\be\label{asymperf}
\lim_{x\to \pm \infty}\operatorname{erf}\bigl(e^{i\theta}x\bigr)=\pm 1
\ee
(e.g., \cite[Sec.~7.1.16]{abramowitz1964handbook}, or note that $|\operatorname{erf}(e^{i\theta}x)-\operatorname{erf}(x\cos\theta)|\leq \tfrac{2}{\sqrt{\pi}}\int_0^{x\sin\theta} dy\, |\exp(-(x\cos\theta+iy)^2)| \leq \tfrac{2}{\sqrt{\pi}} x\sin\theta \exp(-x^2(\cos^2\theta-\sin^2\theta))\to 0$ as $x\to\pm\infty$). Note that $\sigma(\Delta t)$ as in \eqref{sigmaDeltatdef} has phase strictly between $-\tfrac{\pi}{4}$ and $\tfrac{\pi}{4}$ since $\sigma^2>0$. We can now determine the value of $C$ in \eqref{evolvederf} by noting that for from $x=0$, erf is essentially the constant function $\pm 1$, which is stationary under $\exp(-i\hat{H}\Delta t/\hbar)$; thus, $C=0$. (An alternative way of arriving at \eqref{evolvederf} with $C=0$ starts from noting that erf is the convolution $\operatorname{sgn} *\, g$ with $\operatorname{sgn}(x)=x/|x|$ the sign function and $g$ the appropriate Gaussian, so $e^{-i\hat{H}t/\hbar}$ erf = $\operatorname{sgn} * \, g_t$.)

By a Galilean transformation with relative velocity $\hbar k_0/m$ \cite[Exercise 1.18]{Tum22},
\be
e^{-i\hat{H}\Delta t/\hbar}\Bigl[ \operatorname{erf}\Bigl(\frac{x}{2\sigma}\Bigr)\, e^{ik_0x} \Bigr] = \operatorname{erf}\Bigl(\frac{x-\hbar k_0 \Delta t/m}{2\sigma(\Delta t)}\Bigr)\, e^{ik_0x} e^{-i(\hbar k_0^2/2m)\Delta t}\ .
\ee
As a consequence, the evolved wave function takes the form 
\begin{equation}\label{Psintau+erf}
    \Psi(\vx,n\timestep+\Delta t) \approx \frac{\mathcal{N}c}{2} \Biggl[ 1 - \operatorname{erf} \biggl(\frac{|\vx|-\bigl(R + \tfrac{\hbar |\vk_0| \, \Delta t}{m}\bigr)}{2 \sigma_n (\Delta t)} \biggr) \Biggr]e^{i\vk_0\cdot\vx}e^{-i(\hbar \vk^2_0/2m)\Delta t} \ , 
\end{equation}
Since the outward propagation speed is $\hbar |\vk_0|/m=R/n\timestep$, the center of the step has moved to $R+R\Delta t/n\timestep$; the width of the step equals the width of the Gaussian that is the derivative of erf, and that width is 
\be\label{wn}
w_n(\Delta t)=\Bigl(\Re \frac{1}{\sigma(\Delta t)^2} \Bigr)^{-1/2} = \sqrt{\sigma_n^2 + \frac{\hbar^2\Delta t^2}{4m^2 \sigma_n^2}} \,.
\ee

\bigskip

\begin{figure}
    \centering
    \includegraphics[width=10cm, height= 20cm]{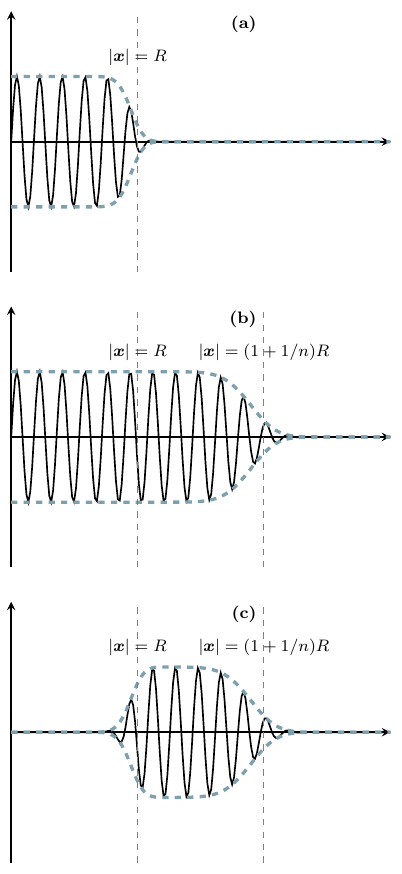}
     \caption{Illustration of $\Psi$ at times (a) $t=n\timestep+$ (no detection), (b) $t=(n+1)\timestep-$, (c) $t=(n+1)\timestep+$ (detection). As in Figure~\ref{softstep}, $\pm |\Psi|$ (dashed) is shown along with $\Re\,\Psi$ (black) in a neighborhood of $\vx_0$ with $|\vx_0|=R$. Between the two measurements, i.e., between (a) and (b), the step moves outward by $R/n$, and its width grows to \eqref{wnRn}.}
    \label{secondapprox}
\end{figure}

\emph{Extent of perturbation inside $\Omega$-} We choose the size of $\sigma_n$ carefully, such that the center of the step moves faster outward than the width of the step grows (so that the step will not affect the wave function inside $\Omega$); that is, we choose $\sigma_n$ such that
\be\label{wnRn}
w_n(\timestep) \ll R/n \,,
\ee
see Figure~\ref{secondapprox}(b). This ensures very little reflection (similar to the case of imaginary potentials, where the reflected wave was of order $\lambda$) in the sense of very little perturbation of $\Psi$ traveling inward. In view of \eqref{wn}, this will be satisfied if both $\sigma_n\ll R/n$ and $\hbar\timestep/2m\sigma_n\ll R/n$; since we take $\sigma_n=n\sigma_1$, the latter condition is equivalent to $\sigma_1\gg \hbar\timestep/mR$ and the former to $\sigma_1\ll R/n^2$; since $n$ measures the time in units of $\timestep$ and relevant times are of order $R$, $R/n^2$ is of order $\timestep^2/R$. Thus, \eqref{wnRn} follows from \eqref{sigmandef}.

As a consequence, the part of the wave function that is following suit from inside the sphere by time $(n+1)\timestep$ around $\vx_0$ is well approximated by a plane wave (Figure~\ref{secondapprox}(b)), which inductively justifies the plane wave assumption in \eqref{Psierfplane}.

\bigskip

\emph{Detection time and place-} Essentially, our calculation of the time evolution of $\hat{P}_{\sigma_n}^{1/2}\Psi$ justifies the first approximation described in Section~\ref{sec:firstapprox}. 
 
In detail, let us compute the probability of detection at the $n^{\mathrm{th}}$ attempt. We have seen that between $n\timestep$ and $(n+1)\timestep$, the step moves outward by $R/n$ and reaches width $w_n(\timestep)\ll R/n$; thus, $\Psi$ leaks out of $\Omega$ at most into a shell of thickness $2R/n$; except for the first few values of $n$ (for which $\Psi$ is still concentrated near the origin), this thickness is small compared to $R$ (like $2/n$), so the normalizing constant $\mathcal{N}$ is dominated by the amount of $\Psi$ still inside $\Omega$, or $\mathcal{N}\approx \mathcal{N}_n$ as in \eqref{calNn}. In particular, the probability of no detection up to $n\timestep$ is given by $\mathcal{N}_n^{-2}$, which means that $T_D/R$ is distributed as in \eqref{cross3}. Since the thickness is small compared to $R$, $|\vX_D|/R\approx 1$ with high probability, see Figure~\ref{secondapprox}(c). Since the step function acts only on the radial coordinate, the angle dependence of $\Psi$ is unchanged and \eqref{cross3} also applies to $\vX_D/|\vX_D|$; thus, \eqref{cross3} applies completely.

An alternative justification can be provided using a Bohmian analysis of the model. Since we already know that the asymptotics of $(T_{WOD},\vX_{WOD})$ is given by \eqref{cross} in the sense of \eqref{cross3}, in order to derive the same asymptotics for $(T_D,\vX_D)$, it suffices to show that the difference is negligible compared to $R$. To this end and to verify \eqref{TDTWIDerrorZeno}--\eqref{XWIDXWODerrorZeno}, let us estimate how small the change on $\Psi$ inside $\Omega$ due to the collapse is: Since $|1-\operatorname{erf}(x)|<\exp(-x^2)$ for sufficiently large real $x$ and $|\operatorname{erf}(e^{i\theta}x)-\operatorname{erf}(x)|<x\exp(-(\text{const.})x^2)$ as shown after \eqref{asymperf}, $|1-\operatorname{erf}(e^{i\theta}x)|<\exp(-\alpha x^2)$ with some $\alpha>0$ for sufficiently large $x$. Thus by time $(n+1)\timestep-$ (i.e., $\Delta t=\timestep-$), since the step has moved outward by $R/n$ and grown in width to $w_n(\timestep)$, the square bracket in \eqref{Psintau+erf} is absolutely smaller for $|\vx|<R$ than $\exp(-\alpha(R/nw_n(\timestep))^2)$. Since $w_n(\timestep)<n\sigma_1 +\hbar\timestep/2mn\sigma_1<2\max\{n\sigma_1,\hbar\timestep/2mn\sigma_1\}$ and $n=\mathcal{O}(R/\timestep)$, we have that $R/nw_n(\timestep) > \min\{\mathcal{O}(\timestep^2/\sigma_1 R), \mathcal{O}(\sigma_1 R/\timestep)\}\gg \mathcal{O}(1)$ by \eqref{sigmandef}, so the square bracket is $\ll \mathcal{O}(1)$, and the perturbation inside $\Omega$ is negligible, which means it can be ignored after the next collapse. Still, at time $n\timestep+$ for (say) $R-\sigma_n \leq |\vx|\leq R$, the perturbation of $\Psi$ is not negligible, with two consequences: $T_D$ might be less than $T_{WID}$ by up to $\timestep$ (i.e., the particle could be detected before it has even reached $S_R$), and $T_{WID}$ might be less than $T_{WOD}$ (as if the particle felt tailwinds): after all, the growth of $w_n(\Delta t)$ entails that probability will be transported outward faster than at speed $|\vv_{WOD}|=\hbar |\vk_0|/m$, at most by the distance $\mathcal{O}(w_n(\timestep))$; since velocities are of order 1 (neither very large nor very small), the particle might arrive earlier by at most $\mathcal{O}(w_n(\timestep))=o(R/n)=o(\timestep)$. In addition, the mere fact that after the particle crosses the sphere $S_R$, the next attempted detection may still occur up to $\timestep$ later can lead to a difference between $T_D$ and $T_{WID}$ of order $\timestep$. We thus obtain \eqref{TDTWIDerrorZeno},
\eqref{XDXWIDerrorZeno} (since velocities are bounded and $\vX_D=\vQ(T_D)$),
\eqref{TWIDTWODerrorZeno}, and (again since velocities are bounded)
\eqref{XWIDXWODerrorZeno}. As a consequence, we obtain \eqref{cross3}.

We still need to verify two properties we asserted in Section~\ref{sec:Zenoresults}: first, that the probability of ``no detection'' at $n\timestep$ agrees with $\langle \Psi_{n\timestep-}|\hat{P}_{\sigma_n}|\Psi_{n\timestep-}\rangle$; and second the equivariance (preservation of the $|\Psi|^2$ distribution) through the collapse.  Here is the calculation for the first property. Consider a measurement at time $n\timestep$. From \eqref{Zdisbn} and the fact that the choice of $Z$ is independent of $\vQ_{WID}(n\timestep-)$, we have that
\begin{align}
    \mathbb{P}\Bigl(\text{``no detection''} \Big| \vQ_{WID}(n\timestep-)=\vq\Bigr)
    &=\PPP \Bigl( |\vQ_{WID}(n\timestep-)|+Z <R \ \Big| \vQ_{WID}(n\timestep-)=\vq\Bigr) \\
    &=\mathbb{P}\bigl(Z<R-|\vq| \bigr) \\
    &=\int_{-\infty}^{R-|\vq|}dz \ p_{\sigma_n}(z)\\
    &= \frac{1}{4}\left( 1-\text{erf}\left(\frac{|\vq|-R}{2\sigma_n} \right)\right)^2. \label{ND|Q}
\end{align}
Therefore,
\begin{align}
\mathbb{P}(\text{``no detection''})&= \int d^3\vq  \ \mathbb{P}\Bigl(\text{``no detection''} \Big| \vQ_{WID}(n\timestep-)=\vq\Bigr) \, \mathbb{P}\bigl(\vQ_{WID}(n\timestep-)=\vq\bigr)\\
&= \frac{1}{4}\int d^3\vq \ \left( 1-\text{erf}\left(\frac{|\vq|-R}{2\sigma_n} \right)\right)^2 |\Psi_{n\timestep-}(\vq)|^2, \\
&= ||\hat{P}^{1/2}_{\sigma_n} \Psi_{n\timestep-}||^2,\label{ND}
\end{align}
which is what we wanted to show. 

We turn to the second property. While the Bohmian trajectories are explicitly equivariant under the continuous Schr\"odinger evolution from time $(n-1)\timestep+$ up to $ n\timestep-$, measurement at time $n\timestep$ necessitates clarification on whether this property holds immediately after, i.e., at time $n\timestep+$. Specifically, we must justify that the distribution of the Bohmian position $\vQ_{WID}(n\timestep+)$, conditionally on either ``no detection'' or ``detection'' at time $n\timestep$, still agrees with $|\Psi_{n\timestep^+}|^2$. Indeed, using the fact that the Bohmian position remains unchanged during the measurement ($\vQ(n\timestep+) = \vQ(n\timestep-)$) as well as \eqref{ND|Q} and \eqref{ND}, we obtain that
\begin{align}
    &\PPP\Bigl(\vQ_{WID}(n\timestep+)=\vq \Big| \text{``no detection''}\Bigr) =  \nonumber\\
    &\qquad = \frac{\PPP\left(\vQ_{WID}(n\timestep+)=\vq, \text{``no detection''}\right)}{\PPP\text{(``no detection'')}} \\
    &\qquad = \frac{\PPP\left( \text{``no detection''}|\vQ_{WID}(n\timestep-)=\vq\right) \, \PPP(\vQ_{WID}(n\timestep-)=\vq)}{\PPP(\text{``no detection''})} \\
    &\qquad = \frac{\frac{1}{4}\left( 1-\text{erf}\left(\frac{|\vq|-R}{2\sigma_n} \right)\right)^2 |\Psi_{n\timestep-}(\vq)|^2}{|| \hat{P}^{1/2}_{\sigma_n}\Psi_{n\timestep-} ||^2} \\
    &\qquad = \frac{|\hat{P}^{1/2}_{\sigma_n}\Psi_{n\timestep-}(\vq)|^2 }{|| \hat{P}^{1/2}_{\sigma_n}\Psi_{n\timestep-} ||^2} =|\Psi_{n\timestep+}(\vq)|^2 . 
\end{align} 
A similar calculation can be done for the event of ``detection'' at time $n\timestep$, which completes the proof of the second property. Furthermore, since in the event of ``detection'' we take for granted that the final position measurement finds the actual position, we have also justified \eqref{PDdefzeno}. 

This concludes our analysis of the Zeno-dynamics-based detector model.

\section{Generalizations}
\label{sec:general}

In this section, we derive and discuss the generalizations of our results to non-spherical shapes, to $N$ particles, moving surfaces, and to the Dirac equation, as summarized in Remarks~\ref{rem:shape}--\ref{rem:Dirac}.

\subsection{Arbitrary Surfaces}
\label{arbsurf}

To extend the analysis to detector surfaces of arbitrary shape, we take the surface to be the boundary $\partial\Omega$ of some region $\Omega\subset \RRR^3$ around the origin. Let $\vn(\vx)$ denote the outward unit normal vector to the surface at $\vx\in\partial \Omega$. As mentioned in Remark~\ref{rem:shape}, we assume that
\be\label{ndotx0}
\vn(\vx)\cdot \vx >0\,.
\ee 
This entails in particular that $\Omega$ is \emph{star-shaped} relative to the origin, i.e.,
\begin{equation}\label{concavesurf}
    \forall \vx \in \Omega,\quad \forall \lambda \in [0,1]: \quad  \lambda\vx  \in \Omega.
\end{equation}
We assume that $\Omega$ is obtained from some set $\Omega_1$ through scaling by a factor $R$, i.e., $\Omega=R\Omega_1$, while the initial wave function $\Psi_0$ is again taken as independent of $R$, so for large $R$ it will be concentrated inside $\Omega$.

Let us first compute the distribution of $T_{WOD}$ and $\vX_{WOD}$. It is known \cite{Daumer,DT09} that the normal component of the current yields the local crossing probability (potentially up to a sign depending on the orientation of the crossing),
\be\label{njWOD}
    \mathbb{P}(\vX_{WOD}\in d^2\vx, T_{WOD} \in dt) = |\vn(\vx)\cdot \vj_{WOD}(\vx,t)| \ d^2\vx \, dt\,.
\ee
From the asymptotic form \eqref{asymppsi} of $\Psi_{WOD}$, it follows that $\vj_{WOD}$ asymptotically points in the outward radial direction, so by \eqref{ndotx0}, $\vn\cdot\vj\geq 0$, and the absolute bars in \eqref{njWOD} can be dropped:
\begin{align}
    \mathbb{P}(\vX_{WOD}\in d^2\vx, T_{WOD} \in dt) 
    &= \tfrac{\hbar}{m}\vn(\vx)\cdot \Im\bigl(\Psi_{WOD}^* \nabla\Psi_{WOD}\bigr) \, d^2\vx \, dt\\
    &\stackrel{\eqref{asymppsi}}{\approx} \frac{m^3\ \vn(\vx)\cdot \vx}{\hbar^3 t^4}\left| \widehat{\Psi}_0\left( \frac{m\vx}{\hbar t}\right)\right|^2  d^2\vx \, dt\,.\label{PPPWODn}
\end{align}

Now consider an imaginary potential $-i\lambda 1_{\Omega^c}(\vx)$. Zooming in on a neighborhood of $\vx\in\partial \Omega$, the surface looks locally like a plane (possibly tilted relative to the radial direction), $\Omega$ looks like a half space, and the incident wave function like a plane wave. Now $\vk$ is not parallel to $\vn$, but the previous calculation of reflected and transmitted amplitudes still applies to the component in the direction of $\vn$. Thus, the reflected wave $\Psi_R$ does not move radially inward, but it still has amplitude $\mathcal{O}(\lambda)$. As in Section~\ref{Impot}, it follows that $\vj_{WID}$ differs from $\vj_{WOD}$ by a $\mathcal{O}(\lambda)$ term, and also $\vv_{WID}-\vv_{WOD} = \mathcal{O}(\lambda)$. Since the travel time from the origin to $\partial \Omega$ is of order $R$, the accumulated corrections to the time and place of arrival at $\partial\Omega$ are of order $\lambda R$:
\be
T_{WID}-T_{WOD} = \mathcal{O}(\lambda R)\,,~~~
\vX_{WID}-\vX_{WOD} = \mathcal{O}(\lambda R), 
\ee
in agreement with \eqref{TWIDTWODerror} and \eqref{XWIDXWODerror}. Since the absorption rate in $\Omega^c$ is $\lambda$, the waiting time until absorption (i.e., detection) is again exponentially distributed with expectation $1/\lambda$, in agreement with \eqref{TDTWIDerror}. Since the velocities are of order 1, also \eqref{XDXWIDerror} remains true. Given that all of these errors are small compared to $R$, the leading-order asymptotics of $\PPP_D$ coincide with those of $\PPP_{WOD}$, i.e., with \eqref{PPPWODn}. This confirms the formula \eqref{crossn}. We remark that this formula can be written as
\begin{align}\label{sigmaarbsurf}
    \sigma(\vx,t)\, d^2\vx\, dt 
    &= \langle \Psi_0| \hat{F}(d^2\vx \times dt) | \Psi_0\rangle,
\end{align}
where $\hat{F}$ is the POVM on $\partial\Omega \times [0,\infty)$ governing the distribution of $(\vX_D,T_D)$, given by
\be\label{Fdef}
\hat{F}(d^2\vx \times dt)= \frac{m^3 \ \vn(\vx) \cdot \vx}{\hbar^3 t^4} |\vk\rangle \langle \vk| \, d^2\vx \, dt
\ee
with $|\vk\rangle$ the momentum eigenstate with wave vector $\vk=m\vx/\hbar t$.

Similar arguments hold when we consider Zeno-type dynamics to model the behavior of detectors: 
a step is created in the wave function through the collapse along the surface $\partial \Omega$; the step gets carried along with the motion in the direction $\vk$ with speed $m|\vk|/\hbar$, so its propagation in the $\vn$ direction occurs with speed $m\vn\cdot\vk/\hbar=\vn\cdot \vx/t$, which is positive and of order 1 (as in the spherical case); the spreading of the step occurs in the direction orthogonal to the step, i.e., $\vn$, so the previous calculations still apply in the direction $\vn$. As a consequence, also according to the Zeno dynamics, the leading-order distribution of $(T_D,\vX_D)$ remains close to that of $(T_{WOD},\vX_{WOD})$ also for non-spherical surfaces. 

Thus, in both models, we arrive at the generalized scattering cross section formula \eqref{crossn}.

\subsection{$N$-Particle Systems}
\label{sec:N}

Now, consider the case where $N$ detectable, non-interacting (but entangled) particles are initially present in the region $\Omega$, and their initial state is given by the joint wave function of the form $\Psi_0(\vx_1, \vx_2, \ldots, \vx_N)$. The framework developed here naturally extends to such multi-particle systems, including the case in which the detector surface is non-spherical (i.e., more general geometries, as described in Section~\ref{arbsurf}). 
Suppose first that the particles are distinguishable. (We will argue below that the same conclusions hold for identical particles.) Let $T_{D,i}\in[0,\infty)$ and $\vX_{D,i}\in\partial\Omega$ denote the time and place of detection of particle $i\in\{1,\ldots,N\}$. As mentioned in Remark~\ref{rem:N}, the definition of the scattering cross section $\sigma$ is in this case
\begin{multline}\label{sigmaNdef}
\sigma(\vx_1,t_1,\ldots,\vx_N,t_N) d^2\vx_1\, dt_1\cdots d^2\vx_N\, dt_N=\\\PPP\Bigl( \vX_{D,1}\in d^2\vx_1,T_{D,1}\in dt_1,\ldots, \vX_{D,N}\in d^2\vx_N, T_{D,N}\in dt_N \Bigr)\,.
\end{multline}
Since the $N$ particles are non-interacting, the experiment can be thought of as $N$ independent detection experiments (one on each particle). It is a general fact of quantum mechanics that the POVM $\hat{F}_N$ of the joint outcomes of experiments on individual particles is the tensor product of the POVMs for each particle. Thus,
\be
\hat{F}_N(d^2\vx_1\times dt_1 \times \cdots \times d^2\vx_N \times dt_N)= \hat{F}(d^2\vx_1 \times dt_1)\otimes \cdots \otimes \hat{F}(d^2\vx_N\times dt_N). 
\ee
By the formula \eqref{Fdef} for $\hat{F}$,
\begin{align} \label{sigmaNparticle}
    &\hspace{-5mm} \PPP\Bigl( \vX_{D,1}\in d^2\vx_1,T_{D,1}\in dt_1,\ldots, \vX_{D,N}\in d^2\vx_N, T_{D,N}\in dt_N \Bigr)\nonumber\\
    &= \langle \Psi_0| \hat{F}_N(d^2\vx_1 \times dt_1 \times \cdots \times d^2\vx_N \times dt_N)| \Psi_0\rangle\\
    &= \frac{1}{\hbar^{3N}}\left( \prod_{i=1}^N \frac{m_i^3\  \vn(\vx_i)\cdot \vx_i}{\ t_i^4}\right)\left| \widehat{\Psi}_0\left(\frac{m_1\vx_1}{\hbar_1t_1}, \ldots, \frac{m_N\vx_N}{\hbar_Nt_N} \right)\right|^2 d^2\vx_1\ldots d^2\vx_N \, dt_1\ldots dt_N\,. 
\end{align}
With \eqref{sigmaNdef}, that yields \eqref{crossnN}. Since this reasoning applies to arbitrary $\Psi_0$ and $m_i$, it also applies if $\Psi_0$ is permutation-symmetric or anti-symmetric and all $m_i$ are equal, and thus for identical particles, leading to a function $\sigma(\vx_1,t_1,\ldots,\vx_N,t_N)$ that is symmetric under permutations of the space-time points $(\vx_i,t_i)$. 

\begin{rem}
    Another way of deriving \eqref{crossnN} is based on, instead of tensor products of observables, collapsing the wave function whenever one particle gets detected (and thus removed from the system under consideration); the appropriate collapse rule, formulated in \cite{detect-several} for the absorbing boundary rule and imaginary potentials, inserts the detected position into the wave function and proceeds with the conditional wave function. In our case, it follows that the joint distribution of the detection events is the same as the one given by $\hat{F}_N$, and thus agrees with \eqref{crossnN}.
\hfill$\diamond$\end{rem}

\begin{rem}
    As in the 1-particle case, the distributions $\PPP_{WOD}$ and $\PPP_{WID}$ agree to leading order with $\PPP_D$, and thus are also given by \eqref{crossnN}. There are two ways of seeing this: First, the formula \eqref{crossnN} was derived for $\PPP_{WOD}$ in \cite{DT04}, and since $T_{WOD} \leq T_{WID} \leq T_D$ (and $\vX_{WID}$ lies between $\vX_{WOD}$ and $\vX_D$ on the Bohmian trajectory), all three distributions have the same asymptotics. Second, for the same reasons as for a single particle and a spherical surface, still $T_{D,i}-T_{WID,i}=\mathcal{O}(1/\lambda)\ll R$ and $T_{WID,i}-T_{WOD,i}=\mathcal{O}(\lambda R) \ll R$ (and velocities are of order 1), so all of the differences are small on the scale of $R$.
\hfill$\diamond$\end{rem}

\begin{rem}
    As shown in \cite{DT04}, the density of $\PPP_{WOD}$ in the far-field regime can also be expressed in terms of a \emph{multi-time wave function} \cite{LPT20}, \cite[sec. 7.4.1]{Tum22}
    \be\label{Phidef}
    \Phi_{WOD}(t_1,\vx_1,\ldots,t_N,\vx_N)
    = e^{i\hbar \nabla_1^2 t_1/2m_1}\cdots e^{i\hbar \nabla_N^2 t_N/2m_N}\Psi_0
    \ee
    (where $\nabla_i$ means the gradient relative to $\vx_i$) according to
    \be\label{sigmaNPhi}
    \sigma_{WOD}(t_1,\vx_1\ldots t_N,\vx_N) = \frac{(\hbar/2i)^N}{\prod m_i} \Phi_{WOD}^* \overleftrightarrow{\partial_1} \cdots \overleftrightarrow{\partial_N} \Phi_{WOD}\,,
    \ee
    where the arrows in $\overleftrightarrow{\partial_i}=\overrightarrow{\partial_i} -\overleftarrow{\partial_i}$ indicate whether the derivative is acting on the $\Phi$ function to the left or the one to the right, and $\partial_i=\vn(\vx_i)\cdot \nabla_i$ is the normal derivative. Since asymptotically $\PPP_D=\PPP_{WOD}$, \eqref{sigmaNPhi} also holds for $\sigma=\sigma_D$. In \cite{detect-dirac}, multi-particle detection probabilities in the near-field regime according to the absorbing boundary rule were expressed in terms of multi-time wave functions. 
\hfill$\diamond$\end{rem}

\begin{rem}
    The case of time-dependent $\Omega$ will be treated for the Dirac equation in Section~\ref{sec:covariantformula} and Appendix~\ref{app:ImDirac}. It can be treated in the same way in the non-relativistic case; for that calculation, one should note that under a Galilean transformation, an imaginary potential $-i\lambda 1_{\Omega^c}(\vx)$ becomes $-i\lambda 1_{\Omega^c}(\vx-\vv t)$. 
\hfill$\diamond$\end{rem}

\subsection{Analog for the Dirac Equation}
\label{sec:Dirac}

For the Dirac equation, we limit our considerations to imaginary potentials and leave out the Zeno dynamics. 
Let us first consider a single particle for which $\Psi=\Psi_{WOD}(\vx,t)$ is $\CCC^4$-valued and follows the free Dirac equation
\be\label{Dirac}
i\hbar \frac{\partial\Psi}{\partial t} = -ic\hbar \valpha\cdot \nabla \Psi + mc^2\beta \Psi\,,
\ee
where $\valpha=(\alpha_1,\alpha_2,\alpha_3)$ is the triple of Dirac alpha matrices, $\beta$ the Dirac beta matrix, and $m>0$ the mass.

\subsubsection{Asymptotics of Dirac Wave Functions}

When Fourier transformed, \eqref{Dirac} reads
\be\label{DiracFourier}
i\hbar\frac{\partial \widehat{\Psi}}{\partial t}(\vk) = M(\vk) \, \widehat{\Psi}(\vk)\,,
\ee
where, in the standard (or Dirac) representation of $\valpha$ and $\beta$,
\be\label{Mkdef}
M(\vk) = \begin{pmatrix}
mc^2 I_2 & \hbar c \vsigma \cdot \vk \\ 
\hbar c \vsigma \cdot \vk & -mc^2 I_2
\end{pmatrix}
\ee
with $I_2$ the $2\times 2$ unit matrix and $\vsigma=(\sigma_1,\sigma_2,\sigma_3)$ the Pauli matrices. It is known (see, e.g., \cite{Tha} or the lemma in Appendix~\ref{app:ImDirac}) that $M(\vk)$ has eigenvalues $\pm\hbar \omega$ with 
\be\label{omegadef}
\omega=\sqrt{c^2\vk^2+m^2c^4/\hbar^2}\,,
\ee
and each eigenvalue has a 2-dimensional eigenspace in $\CCC^4$, the projection to which is given by
\be\label{Pkdef}
P_{\pm}(\vk) = \frac{1}{2} \Bigl(I_4 \pm \frac{1}{\hbar\omega}M(\vk)\Bigr) \,.
\ee

The formula analogous to \eqref{asymppsi} for the asymptotic form of $\Psi=\Psi_{WOD}$ for large $\vx$ and $t$ reads
\be\label{asympDirac}
\Psi(\vx,t) \approx \begin{cases}
0 ~~~\text{if $|\vx| \geq ct$, else} \\
\frac{\hbar\omega^{5/2}}{mc^5} \Bigl((-i t)^{-3/2} P_+(\vk) \widehat{\Psi}_0(\vk) \, e^{i\vk \cdot \vx - i \omega t} + (i t)^{-3/2} P_-(-\vk) \widehat{\Psi}_0(-\vk) \, e^{-i\vk \cdot \vx + i \omega t}\Bigr)
\end{cases}
\ee
with 
\be\label{komegarel}
\vk=\frac{mc\vx}{\hbar \sqrt{c^2t^2-\vx^2}}\,,~~~
\omega=\sqrt{c^2 \vk^2 + m^2 c^4/\hbar^2} = \frac{mc^2/\hbar}{\sqrt{1-\vx^2/c^2 t^2}}~.
\ee
The formula \eqref{asympDirac} expresses that, in the long run, different Fourier modes get separated in space because they move at different velocities (given by $c^2\vk/\omega$, which is the velocity corresponding to relativistic 4-momentum $(\hbar\omega/c,\hbar\vk)$), with one exception: the modes with $k^\mu=(\omega/c,\vk)$ and $-k^\mu=(-\omega/c,-\vk)$ (which has negative energy) \emph{both} move at the \emph{same} velocity $c^2\vk/\omega$. The perhaps surprising phenomenon that two different Fourier modes can move forever at equal velocities is connected to the possibility of positive and negative energies in the Dirac equation. The formula \eqref{asympDirac} will be derived in Appendix~\ref{app:Diracstatphase} using the method of stationary phase. For a $\Psi_0$ with purely positive-energy contributions, the formula simplifies as the $P_-$ term vanishes; for this case, it was proven rigorously in \cite{DP}. Our conclusions are also in line with the mathematical theorems of \cite{Narayanan2025} concerning the long-time asymptotics of Gaussian wave packets and their Bohmian trajectories for the $1$D Dirac equation.

\subsubsection{Covariant Notation}

As an alternative to \eqref{asympDirac}, we will also use the following covariant expression for the asymptotic 1-particle wave function at large $x=x^\mu=(ct,\vx)$:
\be\label{asympcov}
\Psi(x) \approx \begin{cases}
0~~~\text{if $x$ is spacelike or lightlike, else}\\
\frac{mc}{\hbar |x|} \Bigl(e^{i3\pi/4} \widetilde{\Psi}(k) \, e^{-ik_\mu x^\mu} + e^{-i3\pi/4} \widetilde{\Psi}(-k) \, e^{ik_\mu x^\mu} \Bigr),
\end{cases}
\ee
where $|x|=\sqrt{x_\mu x^\mu}$, $k=k^\mu$ is given by
\be\label{kcov}
k^\mu=\frac{mc}{\hbar} \frac{x^\mu}{|x|}\,,
\ee
and $\widetilde{\Psi}$ is the covariant Fourier transform of $\Psi$, defined through the relation
\be\label{tildePsidef}
\Psi(x) = (2\pi)^{-3/2} \int\limits_{-\HHH\cup\HHH} V_3(d^3k) \, e^{-ik_\mu x^\mu}\, \widetilde{\Psi}(k)
\ee
with
\be\label{HHHdef}
\HHH=\{k \in \MMM: k^0>0 \text{ and }|k|=mc/\hbar \}
\ee
the mass shell and $V_3$ again the invariant 3-volume. 

Let us compare the covariant Fourier transform $\widetilde{\Psi}$ to the ordinary Fourier transform at $t=0$ in a given Lorentz frame, $\widehat{\Psi}_0$. In contrast to \eqref{tildePsidef}, the defining relation of the latter is
\be\label{hatPsidef}
\Psi(x) = (2\pi)^{-3/2} \int\limits_{\RRR^3} d^3\vk \, e^{i\vk \cdot \vx} \Bigl(e^{-ik^0 x^0}P_+(\vk) + e^{ik^0 x^0} P_-(\vk)\Bigr) \widehat{\Psi}_0(\vk)
\ee
with
\be\label{k0def}
k^0=(\vk^2+m^2c^2/\hbar^2)^{1/2} \,.
\ee
If we parameterize the mass shell by the space components of the 4-vector $k=k^\mu$, i.e., using the mapping
\be
\pi:\RRR^3 \to \HHH,~~~\pi(\vk) = \bigl(k^0, \vk \bigr)
\ee
with $k^0$ as in \eqref{k0def}, then for any volume element $d^3\vk\subset \RRR^3$, the invariant 3-volume of its lift $\pi(d^3\vk)\subset\HHH$ is
\be
V_3(\pi(d^3\vk)) = \frac{|k|}{k^0} dk^1 \, dk^2 \, dk^3
\ee
with $k=\pi(\vk)$, and \eqref{tildePsidef} can be rewritten as
\be\label{tildePsirewritten}
\Psi(x) = (2\pi)^{-3/2} \int\limits_{\RRR^3} d^3\vk \, \frac{|k|}{k^0}\Bigl( e^{i\vk\cdot\vx} e^{-ik^0 x^0} \widetilde{\Psi}(\pi(\vk)) + e^{-i\vk\cdot\vx} e^{ik^0 x^0} \widetilde{\Psi}(-\pi(\vk)) \Bigr) \,.
\ee
Comparing \eqref{hatPsidef} and \eqref{tildePsirewritten}, we find the conversion formulas
\begin{align}
\widetilde{\Psi}(k) &= \frac{k^0}{|k|} P_+(\vk) \, \widehat{\Psi}_0(\vk)\,,\\
\widetilde{\Psi}(-k) &= \frac{-k^0}{|k|} P_-(-\vk) \, \widehat{\Psi}_0(-\vk)\,,\\
\widehat{\Psi}_0(\vk)&= \frac{|k|}{k^0} \Bigl( \widetilde{\Psi}(\pi(\vk)) + \widetilde{\Psi}(-\pi(-\vk)) \Bigr)\,.
\end{align}
By virtue of these formulas,  \eqref{asympcov} is equivalent to \eqref{asympDirac}.
The generalization to the case of $N$ free particles is straightforward.

\subsubsection{Asymptotic Trajectories}
\label{sec:traj}

The Bohmian equation of motion for a single Dirac particle asserts that the world line is everywhere tangent to the probability current 4-vector field $j^\mu=\overline{\Psi}\gamma^\mu \Psi$. Equivalently,
\be
\frac{d\vX}{dt} = c\frac{\Psi^\dagger \valpha \Psi}{\Psi^\dagger \Psi}(\vX(t))\,.
\ee
While in the non-relativistic case \cite{RDM05} and in the Dirac case with wave functions from the positive-energy subspace of the free Dirac Hamiltonian, the Bohmian world lines of free particles asymptotically become straight lines (i.e., the velocity becomes constant), this is not so for general solutions of the free Dirac equation, as we derive in Appendix~\ref{app:asymptraj}. Rather, the Bohmian world lines asymptotically become \emph{elliptical helices} whose axes are timelike straight lines in direction $k^\mu$, with period length (the time required for one winding in the rest frame of the axis) given by $\pi\hbar/mc^2$ (the Compton wave length divided by $2c$), which for an electron has the value of $4.047 \times 10^{-21}\,\text{s}$, and major radius of the ellipse (spatial size of a winding in the rest frame of the axis) no larger than $\hbar/2mc$ (the Compton wave length divided by $4\pi$), which for an electron is $1.93\times 10^{-13}$ m.
Holland \cite[Sec.~12.3.3]{Hol93} suggested in a similar situation to regard such a periodic motion around a straight line as a Bohmian version of Schr\"odinger's \cite{Schr30} concept of \emph{Zitterbewegung}.\footnote{``Zitterbewegung'' (trembling motion) is a name for the fact that the Heisenberg-evolved position operator for a single Dirac particle has an oscillatory time dependence with the Compton frequency (see, e.g., \cite[Sec.~1.6.1]{Tha}). It seems that Holland was not aware of the general fact that asymptotically as $t\to\infty$ for the free Dirac equation, the trajectories keep oscillating forever (i.e., become helices), a fact that arises from the fact that the Fourier mode with wave vector $k^\mu$ and the one with $-k^\mu$ will never separate in space but will keep jointly influencing the particle's motion. Instead, his example consists of a superposition of two plane waves, one with wave vector $(k^0,\vk)$ with $k^0=(\vk^2+m^2c^2/\hbar^2)^{1/2}$ and the other with $(-k^0,\vk)$, which are two Fourier modes that will ultimately get separated in space for every normalizable wave function, as they move at different velocities $c\vk/k^0 \neq -c\vk/k^0$. As soon as they get separated, the oscillatory motion stops and gets replaced by a straight trajectory; they do not get separated in Holland's example only because the example is not normalizable.}

\subsubsection{No-Signaling Theorem}
\label{sec:nosignaling}

Before deriving \eqref{crossDirac}, we note one of its consequences. One can read off rather directly from \eqref{crossDirac} that in the far-field regime, no signals can be sent between spacelike separated regions by means of moving around detectors---more precisely, that for any Cauchy surface $\Sigma$, the marginal distribution of all detection events on $\partial_{>}{}^4 \Omega \cap \mathrm{past}(\Sigma)$ (which contains all that a ``receiver'' could see) is independent of the shape of $\partial_{>} {}^4\Omega$ in the future of $\Sigma$ (which contains all that a ``sender'' could influence). 

This follows from the fact that the joint distribution of all detection events is the same as the joint distribution of the arrival times and places of $N$ non-interacting \emph{classical} particles starting from the origin (and moving uniformly) with the velocity distribution
\be\label{rhovDirachatPsi}
\rho_v (\vv_1,\ldots,\vv_N) = \frac{1}{\hbar^{3N}} \biggl( \prod_{i=1}^N \frac{m_i^3}{(1-\vv_i^2/c^2)^{5/2}} \biggr) \: \biggl| \widehat{\Psi}_0 \biggl( \frac{m_1 \vv_1}{\hbar \sqrt{1-\vv_1^2/c^2}}, \ldots, \frac{m_N \vv_N}{\hbar \sqrt{1-\vv_N^2/c^2}} \biggr) \biggr|^2 \,.
\ee
Such an ensemble of classical particles obviously cannot be used for superluminal signaling. 

To verify the claim that the cross section would be same we note that, for any $\rho_v$ on $B_c^N$, the scattering cross section on $\partial_{>}{}^4\Omega$ would be
\be\label{sigmaDiracrhov}
\sigma(x_1,\ldots,x_N) = \begin{cases}
0 ~~\text{if any $x_i$ is spacelike or lightlike, else}\\
c^{4N} \biggl( \prod\limits_{i=1}^N \frac{n_\mu(x_i) x_i^\mu}{(x_i^0)^4} \biggr) \: \rho_v\biggl( \frac{c\vx_1}{x_1^0}, \ldots, \frac{c\vx_N}{x_N^0} \biggr) \,.
\end{cases}
\ee
Indeed, for a single particle the probability current of such a classical ensemble at $x\in\MMM_{>}=\{y\in\MMM: y^0>0\}$ would be
\be
j^\mu(x) = \begin{cases}
0 ~~\text{if $x$ is spacelike or lightlike, else}\\
c^3 (x^0)^{-4} \: \rho_v(c\vx/x^0) \: x^\mu
\end{cases}
\ee
because, first, at any time $t>0$, $j^0(ct,\vx) \, d^3\vx$ is the probability of a particle position in $d^3\vx$; second, in the ensemble this probability is $\rho_v(\vv) \, d^3\vv$ with $\vv=\vx/t$ and $d^3\vv=d^3\vx/t^3$; and third, $j^\mu$ is tangent to the trajectory in any ensemble with one trajectory through each $x$. This yields \eqref{sigmaDiracrhov} for $N=1$, and the many-particle version follows by considering first independent random velocities, then convex combinations thereof, and then limits thereof. Now the claim follows from the fact that inserting \eqref{rhovDirachatPsi} into \eqref{sigmaDiracrhov} yields \eqref{crossDirac}.

A discussion of the no-signaling theorem in the near-field regime for the absorbing boundary rule and imaginary potentials will be provided in \cite{PTZT2025}; a special case is studied in \cite{ZTZ2024}.

\subsubsection{Covariant Scattering Cross Section Formula}
\label{sec:covariantformula}

We are now ready to state \eqref{crossDirac} in a form that is covariant and also more general in that it does not presuppose that $\Psi$ has exclusively positive-energy contributions but instead allows for \emph{arbitrary} initial Dirac wave functions $\Psi_0$. It reads as follows:
\be\label{sigma_av}
\sigma(x_1,\ldots,x_N) = \begin{cases}
0~~~\text{if any $x_i$ is spacelike or lightlike, else}\\
\Bigl(\prod\limits_{i=1}^N \frac{m_i^2 c^3}{\hbar^2 |x_i|^2} \Bigr) \!\! \sum\limits_{s_1...s_N=\pm} \!\!\! \overline{\widetilde{\Psi}}(s_1k_1,...,s_N k_N) \bigl[ \nslash(x_1)\otimes \cdots \otimes \nslash(x_N) \bigr] \widetilde{\Psi}(s_1 k_1,..., s_N k_N)
\end{cases}
\ee
with $\nslash(x)=n_\mu(x) \gamma^\mu$ and $k^\mu$ as in \eqref{kcov}. Here, we assume (as in Remark~\ref{rem:Dirac}) that $n_\mu(x)x^\mu>0$, which amounts to ${}^4\Omega$ being star-shaped relative to the origin in $\RRR^4$. As a consequence, also $n_\mu(x) k^\mu(x)>0$, and thus each summand in \eqref{sigma_av} is non-negative.

\begin{rem}
    We have mentioned in Section~\ref{sec:traj} and Appendix~\ref{app:asymptraj} that $j^\mu$ has some quickly oscillating terms, and correspondingly the trajectories are narrow helices around a straight axis. These oscillations are small compared to $R$, so they vanish in the limiting distribution of $T_D/R$ etc.\ as expressed in \eqref{cross3}. (Even on the absolute $\mathcal{O}(1)$ scale they would presumably not be observable in practice because of their small size and period length, even if superpositions of positive and negative energies could be prepared.) That is why we have averaged over these oscillations in \eqref{sigma_av}, which corresponds to dropping the cross term in \eqref{helixv1} and to replacing the helices by their axes.\hfill$\diamond$
\end{rem}

As in the non-relativistic case, the claim is that \eqref{sigma_av} provides the limiting distribution of $(T_D,\vX_D)$ as well as $(T_{WOD},\vX_{WOD})$ and $(T_{WID},\vX_{WID})$. For $(T_{WOD},\vX_{WOD})$, it can be obtained from \eqref{asympcov} by evaluating $j^\mu$, dropping the cross term between positive and negative energy contributions, and noting that it always points outward (so that $n_\mu j^\mu$ is the probability density of arrival). It now remains to show that the differences to $(T_{WID},\vX_{WID})$ and $(T_D,\vX_D)$ vanish in the limit when compared to $R$. Since the $N$-particle case follows from the 1-particle case as in Section~\ref{sec:N}, we focus on the latter.

Much of the reasoning of Section~\ref{Impot} goes through for the Dirac case in a parallel way: Since the detection rate is proportional to $\lambda$, $T_D-T_{WID}=\mathcal{O}(1/\lambda)$ and $\vX_D-\vX_{WID}=\mathcal{O}(1/\lambda)$. We show in Appendix~\ref{app:ImDirac} that also the Dirac equation yields $\refl=\mathcal{O}(\lambda)$ for reflection from the potential step at a timelike part of $\partial_> {}^4\Omega$. Therefore, $\vv_{WID}-\vv_{WOD}=\mathcal{O}(\lambda)$ and thus $T_{WID}-T_{WOD}=\mathcal{O}(\lambda R)$. Since velocities are of order 1, also $\vX_{WID}-\vX_{WOD}=\mathcal{O}(\lambda R)$. Thus, the scattering cross section at the timelike parts of $\partial_> {}^4\Omega$ is given by \eqref{sigma_av}. Concerning the spacelike parts of $\partial_> {}^4\Omega$: Locally, the surface $\partial_> {}^4\Omega$ can be approximated by a spacelike hyperplane; we can Lorentz transform to a frame in which this plane is horizontal; this changes the negative-imaginary scalar potential to an imaginary past-timelike 4-vector potential $A_\mu$ proportional to $\lambda$. Again, the detection rate is proportional to $\lambda$, so the time between crossing the surface and detection is $\mathcal{O}(1/\lambda)$. Since speed is bounded, also the difference in the space coordinates is $\mathcal{O}(1/\lambda)$, and therefore also $T_D-T_{WID}$ and $\vX_D-\vX_{WID}$ in another Lorentz frame. (For other models of detectors on spacelike surfaces, see \cite{LT20,LT22}.) Also in a spacelike region of the surface $\partial_> {}^4 \Omega$, the point $(T_{WID},\vX_{WID})$ of arrival at this surface may differ from $(T_{WOD},\vX_{WOD})$ due to the ``headwinds'' mentioned after \eqref{vdelaydef}, but as above only by an amount of $\mathcal{O}(\lambda R)$. This completes the derivation of \eqref{sigma_av} and \eqref{crossDirac}.

\section{Conclusion}
We have successfully derived the standard scattering cross-section formula using two detector models. In contrast to traditional justifications, we treat detectors as active observers, thereby allowing us to quantify the disturbance due to the detectors. However, in the limit $R \to \infty$, the detector-induced disturbance becomes negligible, so that the flow of probability in the presence of detectors agrees with that in the absence of detectors, which is given by the standard formula \eqref{cross}. Furthermore, generalizations to arbitrary surfaces, multi-particle systems, and Dirac particles are also covered. In the Dirac case, we found that the Bohmian trajectories for the general solutions of the Dirac equation exhibit in the long run a fast, small-scale helical motion (manifesting Zitterbewegung, as suggested by Schr\"odinger in the 1930s). We discuss why absorbing boundary conditions are not applicable here. Collectively, these findings establish a firm foundation for deriving standard scattering formulas from explicit detector models. Some open questions remain for future investigation: Can the considerations presented here be turned into mathematical proofs? Can we derive the detector models from a fundamental, many-body microscopic description of the detector-particle interaction? Can these detector models be shown in general to obey a no-superluminal-signaling theorem \cite{PTZT2025}? Can we quantify how hard or soft real-world detectors are? Can the asymptotics of the incoming \eqref{asymppsi}, reflected \eqref{PsiR}, and transmitted \eqref{PsiT} wavefunctions be rigorously justified from the eigenfunctions of a Hamiltonian in 3 dimensions with a spherically symmetric potential? Lastly, the distribution of detection events in the near-field regime is a topic of active discussion.

\appendix

\section{Stationary Phase Approximation of Free Dirac Wave Function}
\label{app:Diracstatphase}

We set $c=1=\hbar$ and start with the freely evolved wave function $\Psi=\Psi_{WOD}$ written in terms of the Fourier transform as 
\begin{align}
    \Psi(\vx, t)&= (2\pi)^{-3/2} \int_{\RRR^3} d^3\vk\left(P_+(\vk)\widehat{\Psi}_0( \vk)\exp \left(i(\vk\cdot\vx - \omega t)\right) + P_-(-\vk) \widehat{\Psi}_0(-\vk)\exp \left(i(-\vk\cdot\vx + \omega t)\right) \right)\\
    &= (2\pi)^{-3/2} \int_{\RRR^3} d^3\vk\left(P_+(\vk)\widehat{\Psi}_0( \vk)\exp \left(iS_+(\vk)\right) + P_-(-\vk) \widehat{\Psi}_0(-\vk)\exp \left(iS_-(\vk)\right) \right) \label{PsiDstat2}
\end{align}
with $S_\pm(\vk)= \pm\vk\cdot\vx \mp \omega t$. We do the calculation separately for the positive and negative energy contributions. According to the stationary phase reasoning, the only significant contributions to the integral come from $\vk$ values near the $\vk_*$ with $\nabla_{\vk}S_\pm(\vk_*)=0$ because all other will cancel out due to fast oscillation; since
\begin{align}
    \nabla_{\vk}S_\pm (\vk) &= \nabla_{\vk}  \left(\pm \vk \cdot \vx \mp \sqrt{ \vk^2 + m^2} \, t \right) \\
    &= \pm\vx \mp \frac{\vk}{\sqrt{\vk^2 + m^2}} \, t\,,
\end{align}
there is only one $\vk_*$ value for each sign (in fact, the same for both signs), viz.,
\be
\vk_{*}=  \frac{m}{\sqrt{1-\vx^2/t^2}} \frac{\vx}{t}\,.
\ee
Taylor expanding the phase to second order around $\vk_*$ yields, with the abbreviation $\vu=\vk-\vk_{*}$,
\begin{align}
S_\pm(\vk_{*}+\vu)
&\approx S_\pm(\vk_{*}) + \tfrac{1}{2}\sum_{i,j=1}^3 \bigl(\partial_{k_i} \partial_{k_j} S_\pm\bigr) \Big|_{\vk_{*}}u_i u_j\\
&= S_\pm(\vk_*)+ \tfrac{1}{2} \biggl(\mp \frac{t}{\omega_*}\vu^2 \pm \frac{t(\vu\cdot \vk_{*})^2}{\omega_*^3}\biggr)\\
&= S_\pm(\vk_*)\mp \frac{t}{2\omega_*} \Bigl( \vu_\perp^2 + (\underbrace{1-\vk_*^2\omega_*^{-2}}_{=m^2\omega_*^{-2}}) \vu_{||}^2 \Bigr)
\end{align}
with $\vu_\perp$ the part of $\vu= \vu_\perp+ \vu_{||}$ orthogonal to $\vk_*$ and $\vu_{||}$ the part parallel to it.
Substituting this expression back into \eqref{PsiDstat2} yields
\begin{align}
    \Psi(\vx, t) &\approx (2\pi)^{-3/2} \sum_{\pm} \int d^3\vu \, P_\pm(\pm\vk_*\pm\vu) \, \widehat{\Psi}_0( \pm\vk_*\pm\vu) \:\times\nonumber\\
    &\qquad \times \: \exp \bigl(iS_\pm(\vk_{*})\bigr) \, \exp\biggl(\frac{\mp i t}{2\omega_*} \vu_{\perp}^2 \biggr) \, \exp\biggl( \frac{\mp i m^2t}{2\omega_*^3}\vu_{||}^2 \biggr)\\
    &=(2\pi)^{-3/2} \sum_{\pm}\exp \bigl(iS_\pm(\vk_{*})\bigr)  \int_{\vk_*^\perp}d^2\vu_\perp \int_{\RRR\vk_*} d\vu_{||} \, P_\pm(\pm\vk_*\pm\vu_\perp\pm\vu_{||}) \:\times \nonumber\\
    &\quad\quad\quad  \times \: \widehat{\Psi}_0( \pm\vk_*\pm\vu_\perp\pm\vu_{||}) \,  \, \exp\biggl(\frac{\mp i t}{2\omega_*} \vu_{\perp}^2 \biggr) \, \exp\biggl( \frac{\mp i m^2t}{2\omega_*^3}\vu_{||}^2 \biggr)\,.
\end{align}
We change the variables to $\vv_{\perp} = \sqrt{t/2\omega_*} \, \vu_{\perp}$ and $\vv_{||}=m\sqrt{t/2\omega_*^3} \, \vu_{||}$ and obtain:
\begin{align}
    \Psi(\vx, t)
    &\approx(2\pi)^{-3/2} \sum_{\pm}\exp \bigl(iS_\pm(\vk_{*})\bigr)  \int_{\vk_*^\perp}\frac{2\omega_*}{t} \, d^2\vv_\perp \int_{\RRR\vk_*} \frac{1}{m}\sqrt{\frac{2\omega_*^3}{t}} \, d\vv_{||} \:\times\nonumber\\
    &\quad\quad\quad \times \: P_\pm\biggl(\pm\vk_*\pm\underbrace{\sqrt{\frac{2\omega_*}{t}}}_{\to 0 \text{ as }t\to\infty}\vv_\perp\pm\underbrace{\frac{1}{m}\sqrt{\frac{2\omega_*^3}{t}}}_{\to 0}\vv_{||}\biggr) \:\times \nonumber\\
    &\quad\quad\quad  \times \: \widehat{\Psi}_0 \biggl( \pm\vk_*\pm\underbrace{\sqrt{\frac{2\omega_*}{t}}}_{\to 0}\vv_\perp\pm\underbrace{\frac{1}{m}\sqrt{\frac{2\omega_*^3}{t}}}_{\to 0}\vv_{||}\biggr) \,  \, \exp\bigl(\mp i \vv_{\perp}^2 \bigr) \, \exp\bigl( \mp i \vv_{||}^2 \bigr)\,.
\end{align}
Since we are considering the asymptotics as $t\to \infty$ while keeping $\vx/t$ (and thus $\vk_*$) constant, we obtain that
\begin{align}
    \Psi(\vx, t)
    &\approx \frac{\omega_*^{5/2}}{m(\pi t)^{3/2}} \sum_{\pm}\exp \bigl(iS_\pm(\vk_{*})\bigr)  \int_{\vk_*^\perp} d^2\vv_\perp \int_{\RRR\vk_*} d\vv_{||} \, P_\pm(\pm\vk_*) \, 
    \widehat{\Psi}_0(\pm\vk_*) \,  \exp\bigl(\mp i \vv_{\perp}^2 \bigr) \, \exp\bigl( \mp i \vv_{||}^2 \bigr)\\
    &=\frac{\omega_*^{5/2}}{m(\pi t)^{3/2}} \sum_{\pm}\exp \bigl(iS_\pm(\vk_{*})\bigr)    P_\pm(\pm\vk_*)  \widehat{\Psi}_0(\pm\vk_*) \, (\mp i \pi)^{3/2}
\end{align}
using that \cite[Eq.~(9.19)]{DT09}
\be
\int_\RRR dx \, \exp(\pm ix^2)=(\pm i\pi)^{1/2} \,.
\ee
This yields \eqref{asympDirac}.

\section{Derivation of Asymptotic Bohm-Dirac Trajectories}
\label{app:asymptraj}

Here is a derivation (not mathematically rigorous) showing that the asymptotic trajectories are straight lines for positive-energy Dirac wave functions and helices in general. First, we note the algebraic fact that 
\be\label{Ppmalphak}
P_{\pm}(\vk) \valpha P_{\pm}(\vk) = \pm \frac{c\vk}{\omega(\vk)} P_{\pm}(\vk)
\ee
with $\omega$ as in \eqref{komegarel}. This can be verified from \eqref{Pkdef} through direct computation.

For a positive-energy wave function $\Psi_0$, we have that $P_+ (\vk) \widehat{\Psi}_0(\vk) = \widehat{\Psi}_0(\vk)$. Thus, for large $\vx$ and $t$, using \eqref{asympDirac} or \eqref{asympcov} and \eqref{Ppmalphak}, 
\begin{align}
j^\mu(ct,\vx)
&= \Psi^\dagger \gamma^0\gamma^\mu \Psi\\ 
&= \frac{\hbar^2\omega^6}{m^2c^9 t^3} |\widehat{\Psi}_0(\vk)|^2 k^\mu
\end{align}
pointing in the direction $k^\mu$, so that the Bohmian velocity $c\vj/j^0$ depends on $\vx$ and $t$ only through $\vk$ and is equal to $c\vk/k^0=\vk/\omega$. Thus, if the Bohmian particle is in the region governed by a particular Fourier mode at any late time, then the particle moves uniformly at the same velocity as that Fourier mode, and thus follows that Fourier mode. So, $\vX(t)=\frac{c^2\vk}{\omega} t+\vX_0$ (a straight line in direction $k^\mu$) is asymptotically a solution of Bohm's equation of motion, as claimed. The same argument yields the corresponding statement for a negative-energy wave function: a trajectory in the region in which $\Psi$ is a plane wave with wave vector $-k^\mu$ is a straight line in direction $k^\mu$ and thus follows the Fourier mode. 

Now let us turn to a superposition of plane waves with wave vectors $k^\mu$ and $-k^\mu$ as in
\be
\Psi(x) = u_+ e^{-ik_\mu x^\mu} + u_- e^{ik_\mu x^\mu}
\ee
with spinors $u_\pm$, each of which must lie in the appropriate 2d subspace for the plane wave to be a solution of the free Dirac equation. We determine the Bohmian trajectories for this $\Psi$. Since the Bohmian equation of motion for a single particle is Lorentz invariant, we can choose a Lorentz frame convenient for solving it. So perform a Lorentz transformation that carries $k^\mu$ to $(mc/\hbar,0,0,0)$; in the new frame,
\be
\Psi(x) = u_+ e^{-imc x^0/\hbar} + u_- e^{imc x^0/\hbar}
\ee 
with other spinors $u_\pm$ than in the previous equation; now $u_\pm$ lies in the range of $P_\pm(\vzero)$; in particular, $u_+\perp u_-$ in spin space $\CCC^4$. It follows that $\Psi^\dagger \Psi=|u_+|^2 + |u_-|^2$ (with $|\cdot|$ the norm in $\CCC^4$), and the Bohmian equation of motion reduces to
\begin{align}
\frac{d\vX}{dt} 
&= c\frac{(u_+ e^{-i\omega_0t/2} + u_- e^{i\omega_0 t/2})^\dagger \valpha (u_+ e^{-i\omega_0 t/2} + u_- e^{i\omega_0 t/2})}{|u_+|^2 + |u_-|^2}\\
&\!\!\! \stackrel{\eqref{Ppmalphak}}{=} c\frac{|u_+|^2 (\vk/k^0) + |u_-|^2 (\vk/k^0) + 2\,\Re(u_+^\dagger \valpha u_- e^{i\omega_0 t})}{|u_+|^2 + |u_-|^2}\label{helixv1}\\
&\!\!\! \stackrel{\vk=\vzero}{=} \va \cos(\omega_0 t) + \vb \sin(\omega_0 t) \label{helixv}
\end{align}
with $\omega_0=2mc^2/\hbar$ twice the Compton angular frequency and real 3-vectors
\be
\va= \frac{2c\,\Re(u_+^\dagger \valpha u_-)}{|u_+|^2 + |u_-|^2}\,,~~~\vb=-\frac{2c\,\Im(u_+^\dagger \valpha u_-)}{|u_+|^2 + |u_-|^2}\,.
\ee
The general solution of \eqref{helixv} is
\be
\vX(t) = \vX(0) - \frac{\vb}{\omega_0}\cos(\omega_0 t) + \frac{\va}{\omega_0} \sin(\omega_0 t)\,,
\ee
which traces out an ellipse in the plane spanned by $\va$ and $\vb$, centered at $\vX(0)$; the period length is $2\pi/\omega_0=\pi\hbar/mc^2$. The corresponding world lines in space-time are elliptical helices with axis parallel to the $t$ axis through the point $(0,\vX(0))$. Thus, in any Lorentz frame (such as the original frame), the axis points in the direction of $k^\mu$, as claimed. Since every component of $\valpha$ has eigenvalues $\pm 1$ \cite{Tha}, a few lines of algebra using the Cauchy-Schwarz inequality show that $|\va|\leq c$, $|\vb|\leq c$, and $\sqrt{\va^2+\vb^2}\leq c$, so the spatial size of the ellipse is at most $c/\omega_0=\hbar/2mc$.

\section{Reflection Coefficient for Imaginary Potential Step in the Dirac Equation}
\label{app:ImDirac}

In analogy to the computation of the reflection coefficient for the non-relativistic Schr\"odinger equation in Section~\ref{Impot}, we now determine the reflection coefficient for the Dirac equation. We do this in 3 space dimensions because in other dimensions, the type of spin is different, and solutions to the 1d Dirac equation are not necessarily related in a direct way to solutions of the 3d Dirac equation. 

Consider a space-time point $x\in\partial_> {}^4\Omega$ at which the boundary of ${}^4\Omega$ is timelike, and a region around $x$ of a size that is small compared to $R$; in this region, the boundary $\partial_> {}^4\Omega$ can be taken to be a timelike hyperplane. We carry out a Lorentz transformation that makes this hyperplane vertical and maps it to the hyperplane $\{x^3=0\}$. The same Lorentz transformation will map the imaginary scalar potential $-i\lambda$ into an imaginary vector potential $-i\lambda A_\mu$, where $A_\mu$ is a future-timelike vector that is constant in the region considered and has $A_1=0=A_2$.

So in order to determine the magnitude of the reflected wave function, we calculate the reflected wave function for the Hamiltonian
\be\label{HDiracAmu}
\hat{H} = -ic\hbar \valpha\cdot\nabla + \beta mc^2 -i\lambda A_\mu \gamma^0 \gamma^\mu 1_{x^3\geq 0} \,.
\ee
That is, the additional potential applies in the half space $\{x^3\geq 0\}$ and is given there by a constant skew-adjoint matrix $-i\lambda A_\mu \gamma^0 \gamma^\mu= -i\lambda A_0 \, I +i\lambda \vA \cdot \valpha$. As in Section~\ref{Impot}, the key step is to find the eigenfunctions of the operator \eqref{HDiracAmu}. To this end, we need the following statement, which is well known \cite[(1.42)]{Tha} for \emph{real} $\vk$, also for \emph{complex} $\vk$.

\bigskip

\noindent{\bf Lemma.} {\it For complex $\vk\in\CCC^3$ and $m\in\CCC$, the matrix}
\be
M:= c\hbar \valpha \cdot \vk + \beta mc^2
\ee
{\it has complex eigenvalues $\pm\sqrt{\eta}$ with
\be\label{etadef}
\eta := c^2\hbar^2 \vk\cdot \vk + m^2 c^4 \,.
\ee
If $\eta\neq 0$, then each eigenvalue has a 2d eigenspace, and the two eigenspaces are mutually orthogonal if and only if $\Re\,\vk$ is parallel to $\Im \, \vk$. If $\eta=0$ and not both $\vk$ and $m$ vanish, then $M$ is not diagonalizable; its Jordan normal form consists of two $2\times 2$ blocks; in particular, the eigenspace with eigenvalue 0 has dimension 2.}

\bigskip

\noindent{\it Proof.} We write $\va$ and $\vb$ for the real and imaginary parts of $c\hbar \vk$. We first use the known facts $\beta^2=I$, $\{\alpha_i,\beta\}=0$, and $\{\alpha_i,\alpha_j\}=2\delta_{ij} I$ \cite[(1.6)]{Tha} to show that
\begin{align}
M^2 &= (\beta mc^2 + \valpha \cdot [\va +i \vb])(\beta mc^2 + \valpha \cdot [\va +i \vb]) \\
&= \beta^2 m^2c^4 + mc^2 [\va +i \vb] \cdot \{\valpha, \beta\} + \frac{1}{2}\sum_{ij} [a_i+ib_i][a_j+ib_j] \{\alpha_i, \alpha_j\} \\
&= m^2c^4 I + \sum_{ij} \delta_{ij} [a_i+ib_i][a_j+ib_j] I \\
&= (m^2c^4 + \va^2-\vb^2+2i\va\cdot\vb) I= \eta I. \label{was175}
\end{align}
Thus, any eigenvalue must be a root of $\eta$. 

Suppose first $\eta \neq 0$. We show that $M$ is diagonalizable. Since both roots of $\eta$ are non-zero, and since any Jordan block larger than $1\times 1$ with eigenvalue $s\neq 0$ has square different from multiples of $I$ because $(sI + N)^2 = s^2 I + 2sN + N^2$ with $N$ the matrix with ones right above the diagonal and zeroes otherwise, it follows that the Jordan normal form of $M$ must be diagonal; thus, $M$ possesses a basis of eigenvectors. There are two complex roots $\pm s_0$ of $\eta$. 

Each root has a 2d eigenspace: Indeed, $X := \mathrm{tr}[(M-sI)^2] = \mathrm{tr}[M^2 -2sM + s^2] = 4(\eta + s^2)$ since $\mathrm{tr}\, M = 0$. Let $m_\pm$ be the multiplicity of the eigenvalue $\pm s_0$; then $X = m_+(s-s_0)^2 + m_-(s+s_0)^2$. Choose $s=\pm s_0$, then $X = 4m_\mp s_0^2$; by the above, $X=8s_0^2$, so $m_\mp=2$.

It is known \cite[(1.228)]{Tha} that $\va\cdot \valpha$ and $\vb\cdot\valpha$ commute for real vectors $\va,\vb$ if and only if $\va\,||\,\vb$; thus, $M$ is normal (i.e., $MM^\dagger=M^\dagger M$, or the self-adjoint and skew-adjoint parts of $M$ commute) if and only if $\Re\,\vk\,||\,\Im \, \vk$. On the other hand, a diagonalizable matrix $M$ is normal if and only if all eigenspaces are mutually orthogonal.

Now suppose $\eta=0$. Since 0 is the only eigenvalue but $M\neq 0$, $M$ cannot be diagonalizable (unitarily or otherwise). Thus, its Jordan normal form contains at least one block larger than $1\times 1$. No $3\times 3$ or $4\times 4$ Jordan block can occur because the square of such a block is not diagonal (as $N^2\neq 0$), but $M^2$ is a multiple of $I$ by \eqref{was175}. Therefore, $M$ has either two $2\times 2$ Jordan blocks or one $2\times 2$ and two $1\times 1$ blocks. In the latter case, $M$ would have rank 1 (keep in mind that the eigenvalue in all blocks is 0); but we show that $M$ has rank $\geq 2$. If $m\neq 0$, then from inspecting \eqref{Mkdef} (now with complex $m$ and $\vk$) we see that the first two columns are linearly independent due to the $I_2$ factor, so rank $M\geq 2$. Now suppose $m=0$ and $\vk\neq \vzero$. By \eqref{was175}, $\va \cdot \vb= \tfrac{1}{2} \, \Im \, \eta=0$; we rotate our coordinates so that $\va$ points in the $x$-direction and $\vb$ in the $y$-direction. Again by \eqref{was175}, $\va^2 - \vb^2 = \Re \, \eta=0$, so $|\va|=|\vb|$. Since multiplication by a non-zero scalar (such as $|\va|^{-1}$) does not change the rank of $M$, we can assume without loss of generality that $|\va|=1=|\vb|$. Then \eqref{Mkdef} takes the form
\be
M = \begin{pmatrix} 0_2 & \sigma_1 + i\sigma_2 \\ \sigma_1+i\sigma_2 & 0_2 \end{pmatrix}
= \begin{pmatrix} 
0&0&0&2\\
0&0&0&0\\
0&2&0&0\\
0&0&0&0
 \end{pmatrix}\,,
\ee
which has rank 2. It now follows that, when $\eta=0$ and $\vk$ and $m$ do not both vanish, $M$ consists of two $2\times 2$ Jordan blocks.
\hfill$\square$

\bigskip

A basis set of eigenfunctions of $\hat{H}$ in \eqref{HDiracAmu} is given by
\be
\psi(\vx) = \begin{cases}
\psi_{\leq}(\vx):= A e^{i(k^1x^1+k^2x^2+k^3x^3)} + B e^{i(k^1x^1+k^2x^2-k^3x^3)} &\text{for }x^3\leq 0\\
\psi_{\geq}(\vx):= C e^{i(K^1x^1+K^2x^2+K^3x^3)+\lambda A^3x^3/\hbar c} + D e^{i(K^1x^1+K^2x^2-K^3x^3)+\lambda A^3x^3/\hbar c} &\text{for }x^3\geq 0\,,
\end{cases}
\ee
where the spinors $A,B,C,D\in\CCC^4$ have to lie in certain subspaces of $\CCC^4$ in order to be compatible with the eigenvalue $\pm \hbar\omega= \pm \sqrt{\hbar^2 c^2 \vk^2+m^2c^4}$.
The matching conditions
\be\label{matchingDirac}
\psi_{\leq}(x^3=0)=\psi_{\geq}(x^3=0)\,,~~~\partial_3\psi_{\leq}(x^3=0) = \partial_3 \psi_{\geq}(x^3=0)
\ee
require that $K^1=k^1$ and $K^2=k^2$; $K^3\in\CCC$ is determined up to a sign (which can be chosen by exchanging $C$ and $D$) by the condition
\be
\sqrt{\hbar^2 c^2 \vk^2+m^2c^4} = \sqrt{\hbar^2 c^2 \vK^2+m^2c^4}-i\lambda A_0
\ee
that ensures that $\psi_{\leq}$ and $\psi_{\geq}$ are compatible with the same eigenvalue. It follows that for small $\lambda$,
\be\label{K3}
 K^3 = k^3 + i\lambda A_0 \frac{\sqrt{\hbar^2\vk^2 +m^2c^2}}{\hbar^2c k^3} + \mathcal{O}(\lambda^2) 
\ee
As before, we are interested in $D=0$ so, for any given spinor $A$ from the range of $P_\pm(\vk)$, the matching conditions amount to $A+B=C$ and $ik^3(A-B)=(iK^3+\lambda A^3/\hbar c)C$, which leads with \eqref{K3} to
\be
B=\mathcal{O}(\lambda)\,,~~~C=A+\mathcal{O}(\lambda)\,.
\ee
We conclude that the amplitude of the reflected wave is $\mathcal{O}(\lambda)$ also in the Dirac case.

\bigskip
\bigskip

\noindent{\bf Funding.} R.K.\ gratefully acknowledges financial support from German Academic Exchange Service (DAAD) (funding programme/ID: Research Grants - Doctoral Programmes in Germany, 2024/25 
(57693453)). 

\bigskip

\noindent{\bf Acknowledgments.} We thank Lawrence Frolov, Fabian Nolte, and Shadi Tahvildar-Zadeh for helpful discussion.

\end{document}